\documentclass[prd,nofootinbib,english]{revtex4}

\usepackage[utf8]{inputenc}
\usepackage{amsmath}
\usepackage{amsfonts,color}
\usepackage{amssymb,float}
\usepackage{color}
\usepackage{enumitem}
\usepackage{graphicx}
\usepackage{accents}
\usepackage{appendix}
\usepackage{hyperref}
\usepackage{url}
\usepackage{physics}
\usepackage{upgreek}

\hypersetup{
    colorlinks=true,
    linkcolor=blue,
    filecolor=magenta,      
    citecolor=red
}

\newcommand{\udt}[3]{#1^{#2}_{\phantom{#2}#3}}

\newcommand{\dut}[3]{#1_{#2}^{\phantom{#2}#3}}

\newcommand{\lc}[1]{\accentset{\circ}{#1}}
\newcommand{\lo}{\accentset{\circ}{\nabla}}
\newcommand{\pd}[1]{\partial#1} 

\begin{document}

\title{General Effective Field Theory of Teleparallel Gravity}

\author{Maria Mylova}
\email{  mmylova@ewha.ac.kr}
\affiliation{Cosmology, Ewha Womans University, 52 Ewhayeodae-gil, Seoul, Republic of Korea} 

\author{Jackson Levi Said}
\email{jackson.said@um.edu.mt}
\affiliation{Institute of Space Sciences and Astronomy, University of Malta, Malta, MSD 2080}
\affiliation{Department of Physics, University of Malta, Malta}

\author{Emmanuel N. Saridakis}
\email{msaridak@noa.gr}
\affiliation{National Observatory of Athens, Lofos Nymfon, 11852 Athens, 
Greece}
\affiliation{CAS Key Laboratory for Researches in Galaxies and Cosmology, Department of Astronomy, University of Science and Technology of China, Hefei, Anhui 230026, P.R. China.}
\affiliation{Departamento de Matem\'{a}ticas, Universidad Cat\'{o}lica del 
Norte,  Avda. Angamos 0610, Casilla 1280 Antofagasta, Chile}

\begin{abstract}
We construct the Effective Field Theory (EFT) of the teleparallel equivalent of 
general relativity (TEGR).  Firstly, we present the necessary field 
redefinitions of  the scalar field and the tetrads. Then we provide all the 
terms at next-to-leading-order, containing the torsion tensor and its 
derivatives, and derivatives of the scalar field, accompanied by generic 
scalar-field-dependent couplings, where all operators are suppressed by a  scale 
$\Lambda$. Removing all redundant terms using the field redefinitions we result 
to the EFT of TEGR, which includes significantly more terms comparing to the EFT 
of General Relativity. Finally, we present an application in a cosmological 
framework. Interestingly enough, although GR and TEGR are completely equivalent 
at the level of classical equations, we find that their corresponding EFTs   
possess minor but non-zero differences. Hence, we do verify that at higher 
energies the excitation and the features of the extra degrees of freedom are 
slightly different in the two theories, thus making them theoretically 
distinguishable.
Nevertheless, we mention that these differences are suppressed by  the heavy mass 
scale $\Lambda$ and thus it is not guaranteed that they could be measured in 
future experiments and observations.

\end{abstract}

\maketitle

\section{Introduction}

General relativity (GR) has been very successful in describing observations at 
all scales where measurements have been taken. As the basis on which the 
standard cosmological paradigm ($\Lambda$CDM) 
\cite{misner1973gravitation,Clifton:2011jh,Aghanim:2018eyx} is built, it can 
describe the entire evolutionary process and the Universe thermal history up to 
the late-time acceleration \cite{Riess:1998cb,Perlmutter:1998np}. In this  
concordance model  the cosmological constant $\Lambda$ is the source of 
acceleration, while on the other hand, 
cold dark matter (CDM) describes the behaviour of galaxies and galactic scale 
structures. However, despite the numerous successes,   
the $\Lambda$CDM scenario seems to  express several cosmological tensions 
that appear to be growing, related to observations of the 
expansion   and the large scale structure  
\cite{Abdalla:2022yfr,Bernal:2016gxb,DiValentino:2020zio,DiValentino:2021izs,
DiValentino:2020vvd, DES:2017txv,Riess:2021jrx,Brout:2021mpj,Scolnic:2021amr}. 
Additionally,  despite decades of searches, no direct detection of any CDM 
particles has yet occurred \cite{Baudis:2016qwx,Bertone:2004pz}. Finally, at 
the theoretical level the cosmological constant observed value cannot be easily
explained \cite{RevModPhys.61.1,Appleby:2018yci,Ishak:2018his}, while 
General Relativity exhibits the disadvantage of being 
non-renormalizable, and thus it cannot be quantized according to the usual 
techniques \cite{Addazi:2021xuf}.

Having these as motivation, a huge amount of research has been devoted to the 
construction of modifications of gravity, namely gravitational theories that 
possess GR as a particular limit, but being in general renormalizable and/or 
rich enough in order to describe the Universe phenomenomenology more accurately 
\cite{CANTATA:2021ktz}. The usual procedure towards modified gravity
is to start from  General Relativity and extend it in various ways    
\cite{Capozziello:2002rd,Nicolis:2008in}, 
resulting for instance in $f(R)$ gravity \cite{DeFelice:2010aj,Nojiri:2010wj}, Weyl gravity \cite{Mannheim:1988dj},  Horndeski theories \cite{Horndeski:1974wa},
  Gauss-Bonnet gravity \cite{Wheeler:1985nh,Nojiri:2005jg,DeFelice:2008wz}, Lovelock gravity \cite{lovelock1971einstein,Deruelle:1989fj} etc.
  Additionally, in the same framework one may have theories with  different Lagrangians but which generate the same field equations with General Relativity, such as   Unimodular gravity \cite{Unruh:1988in}, $f(R,T)$ gravity \cite{Harko:2011kv}, Rastall gravity \cite{Visser:2017gpz}, and theories with Lagrange multipliers \cite{Capozziello:2010uv}.

However, one can equally well start from the alternative, torsional, 
gravitational formulation and extend it accordingly.
In particular, one can consider that the gravitational interaction is expressed 
through torsion, associated with the teleparallel connection ($\Gamma^\sigma_{\ 
\nu\mu}$)  
\cite{Aldrovandi:2013wha,Cai:2015emx,Krssak:2018ywd,Bahamonde:2021gfp}. This 
connection is curvature-less, namely  it leads to a vanishing teleparallel Ricci 
scalar 
$R(\Gamma^\sigma_{\ \nu\mu}) = 0$, while   the Ricci 
scalar calculated through the 
Levi-Civita connection $\lc{\Gamma}^\alpha_{ \ \mu\nu}$ is not zero, i.e. 
$\lc{R}(\lc{\Gamma}^\alpha_{ \ \mu\nu}) \neq 0$ (the 
over-circle represents objects calculated using the Levi-Civita connection).
Hence, gravity is quantified in the torsion tensor, whose contractions gives 
rise to the torsion scalar $T$ which is equal to the Levi-Civita Ricci scalar 
up to a boundary term. Using $T$ as a Lagrangian leads to exactly the same 
equations with GR and that is why the resulting theory was named   
teleparallel  equivalent of general relativity (TEGR) \cite{Cai:2015emx}.
The interesting feature is that although GR and TEGR are equivalent theories at 
the level of equations, their modifications, for instance $f(R)$ and $f(T)$, 
are not equivalent any more,   corresponding to different modified 
gravity classes. 
 
Since torsional gravity gives rises to novel modified gravity theories, that 
cannot arise in the curvature framework, it has attracted the interest of the 
literature 
\cite{Bahamonde:2021gfp,Ferraro:2006jd,Ferraro:2008ey,
Bengochea:2008gz,Linder:2010py,
Chen:2010va,Zheng:2010am,Geng:2011aj,Bamba:2013jqa,Kofinas:2014owa,
Bahamonde:2019zea, Ualikhanova:2019ygl, 
DavoodSadatian:2019pvq,Bose:2020xdz,Zhao:2022gxl,
Escamilla-Rivera:2021xql,Huang:2022slc,Blixt:2022rpl}. Nevertheless, despite the significant research on torsional 
gravity, the formulation of an  effective field theory (EFT) has not been 
investigated in full detail (see Ref.~\cite{Li:2018ixg}  where  
the EFT approach of torsional gravity around a cosmological background is 
presented, and Ref.~\cite{Casadio:2021zai} where one-loop corrections are explored in the context of teleparallel renormalization). There have also been important works in the general metric-affine approach such as Ref.~\cite{Baldazzi:2021kaf}.

In the present work we are interested in investigating 
the effective field theory of teleparallel gravity around a general 
background. In particular, we desire to probe the
potential differences between the EFT of torsional and curvature gravity.
The covariant formulation for the EFT of curvature gravity 
with a single scalar field was originally studied by Weinberg \cite{2008}, while 
the effective field theory of inflation around a fixed background was studied 
in \cite{Cheung:2007st}.  Thus, it is a very active field of 
research, with important phenomenological applications to inflation, 
scalar-tensor theories, dark energy, gravitational waves and black holes 
\cite{Cabass:2022avo,Tsujikawa:2014mba,Solomon,Maldacena:2011nz,
Crisostomi:2017ugk,Mylova:2019jrj,Mukohyama:2022enj,Hui:2021cpm} (for  
reviews on EFT formulation see 
\cite{petrov,baumann2015inflation,burgess2020introduction}).

Having in mind the above discussion on the structure of GR and TEGR,  it is 
difficult to deduce whether the corresponding EFTs will lead to the same 
dynamics, or whether the next-to-leading  order corrections could lead to 
distinguishable features. Specifically, since the essence of the EFT approach 
is that at very high energies   new degrees of freedom may 
be excited, one does not know a priori whether this will happen in the same way 
in GR and TEGR, although they are equivalent at low energies.
Since this important issue has  not attracted the expected interest in the literature, in this work we   construct  an EFT of torsion gravity and try to address it. 

The manuscript is organised as follows: In Sec.~\ref{sec:eft_intro} we review 
the regular EFT approach in curvature-based gravity. In Sec.~\ref{sec:TEGR} we 
briefly review TEGR and we present the field redefinitions for the tetrads and 
the scalar field. Then in 
Sec.~\ref{sec:build_eft}  we formulate the  EFT of TEGR, while in 
Sec.~\ref{sec:cosmology} we perform an application in a cosmological framework. 
Finally, in Sec.~\ref{sec:conclusion} we summarize our main results and 
conclusions.

\section{The Effective Field Theory approach}
\label{sec:eft_intro}

In this section we briefly review the Effective Field Theory (EFT) approach in 
the usual, curvature, formulation of gravity. 
We follow   the footsteps of Ref.~\cite{2008}, where the most general covariant 
EFT of gravity with a single scalar field was 
considered. For the convenience of the reader, 
we briefly introduce the basic concepts here, as this will be useful in the 
following section when we will clarify the different approach one 
needs to take in constructing a torsional EFT.

The cornerstone behind EFTs is the following: at 
very high energies we expect that new degrees of freedom may be excited. 
Although these heavy particles cannot be probed by current scientific 
experiments, we can still examine their effect at lower energies, using the EFT 
framework. The effect of new physics can be encoded in the low-energy dynamics, 
leading to observational signatures that could be probed by current or next 
generation  experiments, assuming that they are sensitive enough. For 
example, the upper bound on the Hubble parameter during inflation is estimated 
to be at $H_*/m_P < 0.25 \times 10^{-5}$ (95\% \text{CL}), for some pivot scale 
$k_*$ \cite{Planck:2018jri}. Therefore, it is reasonable to assume that 
inflation could be sensitive to higher-energy physics  \cite{2008}. Furthermore, 
 EFTs have also been applied to black holes, where gravitational effects become 
important.

\subsection{EFT of curvature gravity with a  minimally-coupled  scalar field}

In order to construct an EFT for curvature gravity with a  minimally-coupled  
scalar field, one needs to write down the most general Lorentz-invariant 
Lagrangian 
that is a function of the metric, the Riemann tensor and its contractions, and the scalar fields and its derivatives, namely
\begin{equation}
\begin{split}
    \Delta L_{\text{GR},\phi} & \big(g_{\mu\nu},\phi, f(\phi), \pd_\alpha \phi, 
\lc{\square} \phi, \lc{R}, \lc{R}_{\mu\nu}, \lc{R}^\mu_{ \ \nu \rho \sigma} + 
\cdots \big)\,,
\end{split}
\end{equation}
where the Riemann tensor is expressed in terms of the Levi-Civita connection, 
and  $\phi$ is a dimensionless scalar field.

There  are two approaches for the construction of an EFT. If the fundamental (UV 
complete) theory is known, one can take a top-down approach and integrate out 
the heavy degrees of freedom, which are inaccessible to our experiments due to 
their highly-energetic nature. These are particles with mass $M>\Lambda$, where 
$\Lambda$ denotes the cut-off scale at which the EFT breaks down.  One can then 
elaborate the low-energy effective Lagrangian which is described by  higher-dimension operators, suppressed by the heavy mass scale of the new physics 
$\Lambda$. 

On the other hand, if the fundamental theory  is not known, then one can 
equivalently employ Weinberg's theorem and construct the EFT by 
power-counting sub-leading contributions to the zeroth-order Lagrangian. This 
is performed by writing down the most general set of operators consistent with 
the symmetries of the full theory (i.e. Lorentz invariance) and the fundamental 
degrees of freedom (namely the metric $g_{\mu\nu}$  and other particle content 
of the theory, such as the scalar field $\phi$). 

The resulting theory can be expressed  as an energy expansion, using the 
expansion parameter $E/\Lambda$, where $E$ describes the energy of the 
zeroth-order system and $\Lambda$ is the scale of the new physics. 
Schematically, we can write \cite{baumann2015inflation}
 \begin{equation}\begin{split}
	\mathcal{L}_{Eff} = \mathcal{L}[\psi] + \sum c_i 
\frac{O_i[\psi]}{\Lambda^{d_i-4}}\,,  \label{eq:baun1}
\end{split}\end{equation}
with $\mathcal{L}[\psi]$ quantifying the leading-order terms and where the sum 
is understood as a systematic expansion in powers of $E/\Lambda$. The effects 
of the high-energy physics are encoded in the dimensionless couplings $c_i$ of 
the sub-leading operators $O_i$. These operators are usually classed as 
irrelevant, in the sense that their dimension, denoted by $d_i$, is larger than 
$4$. Theoretically, this sum is infinite but in practise it can be truncated at 
the order that matches our current experimental accuracy, making the theory 
finite as far as measurable quantities are concerned.

The leading-order action  for curvature gravity with a single, 
minimally-coupled, scalar field, is written in terms of the Einstein-Hilbert 
action plus the kinetic and potential terms of the  scalar 
field, and reads as
\begin{equation}\begin{split}
	\mathcal{L}_0=  -\sqrt{g} \frac{m_P^2}{2}  \qty[\lc{R} + g^{\mu\nu} \pd_\mu 
\phi \pd_\nu \phi + U(\phi) ]\,, \label{eq:EFT1}
\end{split}\end{equation}
where $R$ is the Ricci scalar, $\phi$ the scalar field with $U$ its
potential, and $m_P \equiv 1/\sqrt{8\pi G}$ is the reduced Planck mass. The 
Lagrangian in (\ref{eq:EFT1}) is considered as the first term in a generic 
effective field theory. By dimensional analysis, we expect the next-to-leading 
order term in the EFT to contain  four spacetime derivatives, with at most two 
acting on the metric and the scalar field at any time, suppressed by inverse 
powers of a heavy mass scale $\Lambda$. Here, for simplicity we take $\Lambda = 
m_P$, and we consider the coefficients of the EFT to be of order unity. Hence, 
the effective Lagrangian which satisfies the above requirements, was found to 
be
\begin{equation}\begin{split}
	\Delta \mathcal{L} & =  \sqrt{g} \Big[ f_1(\phi) \qty(g^{\mu\nu}  
\phi_{,\mu} \phi_{,\nu})^2 + f_2(\phi) g^{\rho\sigma} \phi_{,\rho} 
\phi_{,\sigma} \square \phi + f_3(\phi) \qty(\square \phi)^2 + f_4(\phi) 
R^{\mu\nu} \phi_{,\mu} \phi_{,\nu} + f_5(\phi) R g^{\mu\nu} \phi_{,\mu} 
\phi_{,\nu}
	\\& + f_6(\phi) R \square \phi + f_7(\phi) R^2 + f_8(\phi) R^{\mu\nu}  
R_{\mu\nu} + f_9(\phi) R^{\mu\nu\rho\sigma} R_{\mu\nu\rho\sigma} + f_{10}(\phi) 
\epsilon^{\mu\nu\rho\sigma} R_{\mu\nu}^{ \ \ \kappa\lambda} 
R_{\rho\sigma\kappa\lambda} \Big]\,,\label{eq:EFT2}
\end{split}\end{equation}
where commas denote partial derivatives and $\epsilon^{\mu\nu\rho\sigma}$ is  
the totally antisymmetric tensor density.

\subsection{Exorcising ghosts}
\label{exo}

The higher-order operators in Eq.~(\ref{eq:EFT2}) may  contain time-derivatives 
that lead to higher than second-order equations of motion,  known as   
Ostrogradski instability or Ostrogradski ghosts 
\cite{ostrogradsky1850memoire,Woodard:2015zca}. This can be problematic, 
since in the absence of extra constraints or symmetries, the Hamiltonian will 
be, almost always, unbounded from below. Fortunately, due to the suppressed 
nature of higher-order operators, characterised by powers of a small expansion 
parameter $(H/\Lambda)$, these can be dealt perturbatively, order-by-order, in 
the energy expansion \cite{georgi1991shell,Grosse-Knetter:1993tae,Arzt:1993gz} 
(see also \cite{Manohar:2018aog} for a pedagogical treatment). A widely used 
technique is to remove unwanted operators by performing a local field 
redefinition. This shifts the action so that problematic operators are pushed to 
even higher-orders in the perturbative expansion, and therefore they become 
redundant. 

We would like to emphasize here  that the reason these runaway instabilities 
cannot be excited in an EFT is because the mass of the Ostrogradski ghost is the 
order of the scale of the new physics $\Lambda$. Thus, they are considered 
to be unphysical, as such energy scales are beyond the validity of the EFT and 
cannot be excited by the low-energy degrees of freedom. 

A particularly popular  technique used for exorcising ghosts, is to perform 
field-redefinitions by repeated application of the equations of motion. This can 
be done by utilising the leading-order Einstein field equations and Klein-Gordon 
equation, given by
\begin{eqnarray} 
\label{eq:EFT3a}
    &&R_{\mu\nu} = - \qty[\phi_{,\mu} \phi_{,\nu} + U(\phi) g_{\mu\nu}],  \\
    &&
\square \phi = U^\prime(\phi)\,.\label{eq:EFT3}
 \end{eqnarray}
Let us consider an explicit example. Using (\ref{eq:EFT3a}),(\ref{eq:EFT3}) the 
term proportional to $f_8$  in (\ref	{eq:EFT2}) can be expressed as
\begin{equation}
    \begin{split}
    f_8 (\phi) R^{\mu\nu} R_{\mu\nu} =  \qty[\phi_{,\mu}  \phi_{,\nu} + U(\phi) 
g_{\mu\nu}]^2\,.\label{eq:EFT4}
\end{split}
\end{equation}
It is now easy to see that this term provides corrections to the kinetic and 
potential  terms in (\ref{eq:EFT1}), which can be absorbed by a redefinition of 
the scalar field, the potential and the $f_1(\phi)$ term.
	
Repeated application of the equations (\ref{eq:EFT3a}),(\ref{eq:EFT3}), can 
assist to rewrite the terms $R$, $R_{\mu\nu}$ and $\square \phi$, in the 
effective Lagrangian (\ref{eq:EFT2}),
in terms of single derivatives of the 
scalar field and corrections to the potential (note that the terms proportional 
to $f_9$ and $f_{10}$ cannot be removed by field redefinitions).
This is an important step as it puts the $f_1$-$f_8$ terms in the effective 
Lagrangian into a much simpler form that can easily be dealt by further  
redefinitions of the scalar field, the potential and the $f_1(\phi)$ term, and 
in this way one avoids having to consider field redefinitions of the metric, 
where the calculations are much more involved.

\section{Teleparallel equivalent of general relativity and field redefinitions}
\label{sec:TEGR}

In this section we review briefly the teleparallel equivalent of general 
relativity (TEGR). Additionally, we present the necessary field redefinitions 
that are going to be used in the EFT construction of the next section.
As we will see there,  the  simple technique described in the 
case of curvature gravity above, in order to bring   the terms in the 
effective Lagrangian into a much simpler form that can   be handled by 
scalar-field redefinitions without the need for metric redefinitions,
does not work for the torsional EFT. In particular, the equations of motion in 
TEGR are not helpful, unless one chooses a specific background (as it was 
done in \cite{Li:2018ixg}), in which case the various terms can be grouped and 
simplified. As our ambition in this work is to construct the most general 
covariant torsional EFT with a  minimally-coupled  scalar field, apart from 
scalar-field redefinition we will need  field redefinitions of the 
tetrad too.   
 
\subsection{Teleparallel equivalent of general relativity }

Let us first present briefly the  TEGR. Teleparallel gravity describes gravity 
in a torsional geometric framework 
\cite{Aldrovandi:2013wha,Cai:2015emx,Krssak:2018ywd}, which is formulated by an 
exchange of the Levi-Civita connection $\udt{\lc{\Gamma}}{\sigma}{\mu\nu}$ 
(over-circles  denote quantities calculated using the Levi-Civita connection) 
with the   teleparallel connection $\udt{\Gamma}{\sigma}{\mu\nu}$.    This 
change of connections implies that the teleparallel Riemann tensor will vanish 
since the teleparallel connection is curvature-less.

In the case of curvature gravity one uses the metric tensor $g_{\mu\nu}$ as 
the fundamental dynamical variable, whereas in torsional gravity one uses the 
tetrad $\udt{e}{a}{\mu}$ (and its inverses $\dut{E}{A}{\mu}$) and the spin 
connection $\udt{\omega}{A}{B\mu}$, which is a flat connection responsible for 
incorporating the local Lorentz invariance of the eventual theoretical setting 
 \cite{Bahamonde:2021gfp}. The tetrad,
namely the orthonormal vectors on the tangent space, can be used to determine 
the metric through
\begin{align}\label{metric_tetrad_rel}
    g_{\mu\nu}=\udt{e}{A}{\mu}\udt{e}{B}{\nu}\eta_{AB}\,,& &\eta_{AB} =  
\dut{E}{A}{\mu}\dut{E}{B}{\nu}g_{\mu\nu}\,,
\end{align}
where Latin indices represent coordinates  on the tangent space and Greek 
indices represent coordinates on the general manifold \cite{Cai:2015emx}. 
  For internal consistency, tetrads  
satisfy
\begin{align}
    \udt{e}{A}{\mu}\dut{E}{B}{\mu}=\delta^A_B\,,&  
&\udt{e}{A}{\mu}\dut{E}{A}{\nu}=\delta^{\nu}_{\mu}\,.
\end{align}
 Finally, the teleparallel connection is defined as 
\cite{Weitzenbock1923,Krssak:2018ywd}
\begin{equation}\label{eq:tg_connection}
    \udt{\Gamma}{\sigma}{\nu\mu} := \dut{E}{A}{\sigma} 
\left(\partial_{\mu}\udt{e}{A}{\nu} + 
\udt{\omega}{A}{B\mu}\udt{e}{B}{\nu}\right)\,.
\end{equation}
The tetrad and spin connection components represent the gravitational  and local 
Lorentz degrees of freedom of the equations of motion   
\cite{Hayashi:1979qx,Aldrovandi:2013wha}. The vanishing of 
the Riemann tensor implies that a new measure of geometric deformation is 
needed to formulate gravitational interactions. The case where the spin 
connection components are compatible with zero values is called the 
Weitzenb\"{o}ck gauge. The torsion tensor  reads as \cite{Hayashi:1979qx}
\begin{equation}
    \udt{T}{\sigma}{\mu\nu} := 2\udt{\Gamma}{\sigma}{[\nu\mu]}\,,
\end{equation}
with square brackets denoting an antisymmetric operator, and where 
$\udt{T}{\sigma}{\mu\nu}$ represents the gauge field strength of gravity in TEGR 
\cite{Aldrovandi:2013wha}. This is covariant under both diffeomorphisms and 
local Lorentz transformations, and by taking suitable contractions leads to the 
 torsion scalar 
\cite{Krssak:2018ywd,Cai:2015emx,Aldrovandi:2013wha,Bahamonde:2021gfp}
\begin{equation}
    T := \frac{1}{4}\udt{T}{\alpha}{\mu\nu}\dut{T}{\alpha}{\mu\nu} +  
\frac{1}{2}\udt{T}{\alpha}{\mu\nu}\udt{T}{\nu\mu}{\alpha} - 
\udt{T}{\alpha}{\mu\alpha}\udt{T}{\beta\mu}{\beta}\,.
\label{torsionscalar}
\end{equation}
We mention that   the 
curvature-based Ricci scalar $\lc{R}$ is equal to the torsion scalar plus  a 
total divergence term, namely 
\cite{Bahamonde:2015zma,Farrugia:2016qqe}
\begin{equation}\label{LC_TG_conn}
    R=\lc{R} + T - B = 0\,,
\end{equation}
where $B = (2/e) \partial _\rho (eT^\mu {}_\mu{}^\rho)$
is the boundary term, and $e = \det\left(\udt{e}{A}{\mu}\right) = \sqrt{-g}$ 
is the determinant of the tetrad. 

Since $\lc{R}$ and $T$ differ by a boundary terms, the theory that uses 
$\lc{R}$ as a Lagrangian, i.e. GR, will be completely equivalent to the theory 
that uses $T$ as a Lagrangian, namely TEGR. In particular,
 the TEGR action is written as
\begin{equation}
    S = -\frac{1}{2\kappa^2}\int T\,e\,\dd^4x + \int \mathcal{L}_{\rm m} e \,\dd^4 x\,,
\end{equation}
where $\kappa^2=8\pi G$ and where $\mathcal{L}_{\rm m}$ is the matter 
Lagrangian. By taking a variation of this action with respect to the 
tetrad, the field equations turn out to be 
\begin{equation}
    \frac{1}{\kappa^2} \left[\frac{1}{e}\partial 
_{\sigma}(e\dut{S}{A}{\mu\sigma})-\udt{T}{\sigma}{\nu A} \dut{S}{\sigma}{\nu\mu} 
+ \frac{1}{2}\dut{E}{A}{\mu} T + \udt{\omega}{B}{A \nu}\dut{S}{B}{\nu\mu}\right] 
= \dut{\Theta}{A}{\mu}\,,
\end{equation}
where 
\begin{equation}
    \dut{\Theta}{A}{\mu} = \frac{1}{e}\frac{\delta (e 
\mathcal{L}_{\text{m}})}{\delta \udt{e}{A}{\mu}}\,,
\end{equation}
is the energy-momentum tensor, while if the variation is taken with respect to the 
spin connection the corresponding equations of motion vanish 
identically for TEGR. Finally, we can directly add a minimally-coupled, 
canonical scalar field $\phi$ to this action, namely
\begin{equation}
    S_{\rm TEGR}+S_\phi = \frac{1}{2\kappa^2} \int \dd^4{x}  \left\{-e T+ e 
\left[\lc{\nabla}_\mu \phi \lc{\nabla}^\mu \phi - V(\phi)\right]\right\}\,.
\label{eq:lotg}
\end{equation}
This will be the leading-order action of the effective field theory presented 
below.

\subsection{Field redefinitions of the tetrad and  torsion 
scalar}\label{sec:redef_of_tg}

Since we have presented the basics of TEGR, in this subsection we proceed to 
the field redefinition of the tetrads and the torsion scalar, following the 
methods of \cite{Zumalacarregui:2013pma,Solomon}.
Let us   define
\begin{equation}
    \tilde{e}^A_{ \ \mu} = e^A_{ \ \mu} + \delta e^A_{ \ 
\mu}\,,\label{eq:redef4b}
\end{equation}
where $\delta e^A_{ \ \mu} \ll 1$ are   small perturbations, and we 
keep terms up to order $\mathcal{O}(\delta)$, i.e. terms 
$\mathcal{O}(\delta^2)$ and higher order    
are suppressed. 
Consequently, we have
\begin{equation}
    \tilde{g}^{\alpha\rho} = \eta^{AB} \tilde{E}_A^{ \ \alpha} \tilde{E}_B^{ \ 
\rho} = g^{\alpha\rho} + \eta^{AB} \qty( E_A^{ \ \alpha} \delta E_B^{ \ \rho} + 
E_B^{ \ \rho} \delta E_A^{ \ \alpha} )\,.\label{eq:redef4a}
\end{equation} 
Additionally, since the teleparallel 
connection is defined  in \eqref{eq:tg_connection},  for any two 
tetrad-spin connection pairs we have
\begin{equation}
    \mathcal{K}^\alpha_{ \ \mu\nu} \equiv \tilde\Gamma^\alpha_{ \ \mu\nu} - 
\Gamma^\alpha_{ \ \mu\nu} = \tilde{E}_A^{ \ \alpha} \nabla_\nu \tilde{e}^A_{ \ 
\mu} = E_A^{ \ \alpha} \nabla_\nu \delta e^A_{ \ \mu} + \mathcal{O}(\delta^2) = 
\nabla_\nu \delta e^\alpha_{ \ \mu} + \mathcal{O}(\delta^2)\,,\label{eq:redef2}
\end{equation} 
which implies covariance of     $\mathcal{K}^\alpha_{ \ \mu\nu}$. 
From this  it follows that  
\begin{equation}
    \tilde{T}^\sigma_{ \ \mu\nu} = \tilde\Gamma^\sigma_{ \ \mu\nu} - \tilde\Gamma^\sigma_{ \ \nu\mu} = T^\sigma_{ \ \mu\nu} + \mathcal{K}^\sigma_{ \ \mu\nu} - \mathcal{K}^\sigma_{ \ \nu\mu} = T^\sigma_{ \ \mu\nu} + 2 \mathcal{K}^\sigma_{ \ [\mu\nu]} =T^\sigma_{ \ \mu\nu} + 2 \tilde{E}_A^{ \ \sigma} \nabla_{[\nu} \tilde{e}^A_{ \ \mu]}\,,\label{eq:redef3}
\end{equation} 
while  contracting the torsion tensor   gives
\begin{equation}
    \tilde{T}_\mu = \tilde{T}^\sigma_{ \ \sigma\mu} =T_\mu + 2 \mathcal{K}^\sigma_{ \ [\sigma\nu]} = T_\mu + 2 \tilde{E}_A^{ \ \sigma} \nabla_{[\mu} \tilde{e}^A_{ \ \sigma]}\,.\label{eq:redef4}
\end{equation} 
We can use  the above relations to extract an expression for the torsion scalar 
$\tilde{T}$. Calculating the individual terms we find
\begin{equation}
 \tilde{T}_\mu \tilde{T}^\mu = \tilde{g}^{\alpha\mu} \tilde{T}_\mu \tilde{T}_\alpha = T^\mu T_\mu + 2 T^\alpha T^\mu \delta e_{\alpha\mu} + 4 T^\mu \nabla_{[\mu} \delta e_{\sigma ]}^{ \ \ \sigma}\,,\label{eq:redef6}
\end{equation} 
\begin{align}
    \tilde{T}^{\rho\sigma\mu} \tilde{T}_{\rho\sigma\mu}  & =    \tilde{g}^{\rho\alpha} \tilde{g}^{\sigma\beta}\tilde{g}^{\mu\lambda} \tilde{T}_{\rho\sigma\mu} \tilde{T}_{\alpha\beta\lambda}
    = T^{\rho\sigma\mu} T_{\rho\sigma\mu} + 4\, T^{\rho\sigma\mu}   \mathcal{K}_{ \rho [\sigma\mu]} + 2 \delta e^{\rho}_{ \ \alpha} \qty(T_{ \rho \sigma \mu} T^{\alpha\sigma\mu} + T_{ \beta \ \mu}^{ \ \alpha} T^{\beta\ \mu}_{ \ \rho} + T_{\lambda\sigma}^{ \ \ \alpha} T^{\lambda \sigma}_{ \ \ \rho})\nonumber\\
    & = T^{\rho\sigma\mu} T_{\rho\sigma\mu} +  4\, T^{\rho\sigma\mu} \nabla_{[ \mu} \delta e_{\sigma ] \rho}  + 2 \delta e^{\rho}_{ \ \alpha} \qty(T_{\rho\mu\nu} T^{ \alpha \mu\nu} + 2 T_{\mu\nu}^{ \ \ \alpha} T^{\mu\nu}_{ \ \ \rho})\,,\label{eq:redef7}
\end{align} 
\begin{align}
    \tilde{T}^{\rho\sigma\mu} \tilde{T}_{\sigma\rho\mu}  & =    \tilde{g}^{\rho\alpha} \tilde{g}^{\sigma\beta}\tilde{g}^{\mu\lambda} \tilde{T}_{\sigma\rho\mu} \tilde{T}_{\alpha\beta\lambda}\nonumber\\
    & = T^{\rho\sigma\mu} T_{\sigma\rho\mu} + 2\, T^{\rho\sigma\mu}   \mathcal{K}_{ \sigma [\rho\mu]} +  2\, T^{\sigma\rho\mu}   \mathcal{K}_{ \rho [\sigma\mu]} + 2 \delta e^{\rho}_{ \ \alpha} \qty(T_{ \sigma\rho\mu} T^{\alpha\sigma\mu} + T_{ \ \beta  \mu}^{  \alpha} T^{\beta\ \mu}_{ \ \rho} + T_{\sigma\lambda}^{ \ \ \alpha} T^{\lambda \sigma}_{ \ \ \rho})\nonumber\\
    & =  T^{\rho\sigma\mu} T_{\sigma\rho\mu} + 4\, T^{\sigma\rho\mu} \nabla_{[ \mu} \delta e_{\sigma ] \rho}  + 2 \delta e^{\rho}_{ \ \alpha} \qty(T_{\mu\nu}^{ \ \ \alpha} T^{\nu\mu}_{ \ \ \rho} + 2 T_{\mu\rho\nu} T^{\alpha\mu\nu})\,.\label{eq:redef8}
\end{align} 
Hence, assembling them into (\ref{torsionscalar}) we find that 
  the torsion scalar transforms as
\begin{align}
    \tilde{T} & = - \frac{1}{2} \tilde{T}_\mu \tilde{T}^\mu + \frac{1}{8} \tilde{T}^{\rho\sigma\mu} \tilde{T}_{\rho\sigma\mu}  + \frac{1}{4} \tilde{T}^{\rho\sigma\mu} \tilde{T}_{\sigma\rho\mu}  = - \frac{1}{2} T_\mu T^\mu + \frac{1}{8} T^{\rho\sigma\mu} T_{\rho\sigma\mu}  + \frac{1}{4} T^{\rho\sigma\mu} T_{\sigma\rho\mu}\nonumber\\
    & + \delta e^\rho_{ \ \alpha} \qty[- T^\alpha T_\rho + \frac{1}{4} \qty(T_{\rho\mu\nu} T^{ \alpha \mu\nu} + 2 T_{\mu\nu}^{ \ \ \alpha} T^{\mu\nu}_{ \ \ \rho}) + \frac{1}{2} \qty(T_{\mu\nu}^{ \ \ \alpha} T^{\nu\mu}_{ \ \ \rho} + 2 T_{\mu\rho\nu} T^{\alpha\mu\nu})]\nonumber\\
    & + \nabla_{[ \mu} \delta e_{\sigma ] \rho} \qty(- 2 g^{\rho\sigma} T^\mu + \frac{1}{2} T^{\rho\sigma\mu} + T^{\sigma\rho\mu})\,.\label{eq:redef9}
\end{align} 
Additionally, since \eqref{eq:redef4b} can be written as $\tilde{e}^A_{ \ \mu} = 
e^A_{ \ \mu} + e^A_{ \ \sigma} \delta e^A_{ \ \mu}$, the determinant transforms 
as
\begin{equation}
    \det \tilde{e}^A_{ \ \mu}   = \det e^A_\sigma \det\qty(\delta^\sigma_\mu+ \delta e^\sigma_{ \mu}) = e \qty(1 + \delta e^\mu_{\ \mu})\,.\label{eq:redef10}
\end{equation} 
Hence, we find that the leading-order Lagrangian  transforms as
\begin{align}
    \tilde{e} \tilde{T} & =  e \bigg\{ \qty( - \frac{1}{2} T_\mu T^\mu + \frac{1}{8} T^{\rho\sigma\mu} T_{\rho\sigma\mu}  + \frac{1}{4} T^{\rho\sigma\mu} T_{\sigma\rho\mu} ) + \delta e^\beta_{\ \beta} \qty( - \frac{1}{2} T_\mu T^\mu + \frac{1}{8} T^{\rho\sigma\mu} T_{\rho\sigma\mu}  + \frac{1}{4} T^{\rho\sigma\mu} T_{\sigma\rho\mu} )\nonumber\\
    & + \delta e^\rho_{ \ \alpha} \qty[- T^\alpha T_\rho + \frac{1}{4} \qty(T_{\rho\mu\nu} T^{ \alpha \mu\nu} + 2 T_{\mu\nu}^{ \ \ \alpha} T^{\mu\nu}_{ \ \ \rho}) + \frac{1}{2} \qty(T_{\mu\nu}^{ \ \ \alpha} T^{\nu\mu}_{ \ \ \rho} + 2 T_{\mu\rho\nu} T^{\alpha\mu\nu})]\nonumber\\
    & + \nabla_{[ \mu} \delta e_{\sigma ] \rho} \qty(- 2 g^{\rho\sigma} T^\mu + \frac{1}{2} T^{\rho\sigma\mu} + T^{\sigma\rho\mu}) \bigg\}\,.\label{eq:redef11}
\end{align} 
In the following, for convenience we identify $E_{A\mu} \delta e^A_{ \ \nu} 
=\delta e_{\mu\nu}$ with a generic non-symmetric operator 
$L_{\mu\nu}$ (this occurs due to fact that $\delta e _\alpha^\beta$ has 
no natural symmetries). Thus, we have
\begin{align}
    \tilde{e} \tilde{T} & =  e \bigg\{  T + L^\beta_{\ \beta} T + \nabla_{[ \mu} 
L_{\sigma ] \rho} \qty(- 2 g^{\rho\sigma} T^\mu + \frac{1}{2} T^{\rho\sigma\mu} 
+ T^{\sigma\rho\mu})\nonumber\\
    & + L^\rho_{ \ \alpha} \qty[- T^\alpha T_\rho + \frac{1}{4}  
\qty(T_{\rho\mu\nu} T^{ \alpha \mu\nu} + 2 T_{\mu\nu}^{ \ \ \alpha} T^{\mu\nu}_{ 
\ \ \rho}) + \frac{1}{2} \qty(T_{\mu\nu}^{ \ \ \alpha} T^{\nu\mu}_{ \ \ \rho} + 
2 T_{\mu\rho\nu} T^{\alpha\mu\nu})]\bigg\}\,.\label{eq:redef11a}
\end{align} 
Keeping in mind that $L_{\mu\nu}$ is a non-symmetric  operator, we can use the 
following ansatzes to make field redefinitions:
\begin{align}
    &  L_{\mu\nu} = a_1(\phi) \, T_\mu \nabla_\nu \phi\,,\label{eq:L1}\\
  & L_{\mu\nu} = a_2(\phi) \, T_{\mu\nu\beta} \nabla^\beta 
\phi\,,\label{eq:L2}\\
    &  L_{\mu\nu} = a_3(\phi) \, T_{\mu\nu\beta} T^\beta\,, \label{eq:L3}\\
    & L_{\mu\nu} = a_4(\phi) \, \nabla_\mu T_\nu\,,\label{eq:L4}\\ 
    & L_{\mu\nu} = a_5(\phi) \, \nabla^\beta T_{\mu\nu\beta}\,, \label{eq:L5}\\
    & L_{\mu\nu} =  a_6(\phi) \, \nabla_\mu \nabla_\nu \phi\,, \label{eq:L6}\\
    & L_{\mu\nu} =  a_7(\phi) \, T_{\mu}^{ \ \alpha\beta} T_{\alpha\nu\beta}\,,  
\label{eq:L7}
\end{align}
where we have omitted that the coefficients $a_n$ contain an inverse square  
heavy mass scale $\Lambda$, characterising the energy scale of the new physics 
(one can think of the coefficients $a_n$ as having dimension of 
inverse mass squared). We mention here that these are the only operators we can 
define for field redefinitions in agreement with the symmetries of $L_{\mu\nu}$. 
For instance, terms of the form $T_\nu \nabla_\mu \phi, T_{\nu\mu\beta} 
\nabla^\beta \phi, T_{\nu\mu\beta} \nabla^\beta \phi$ and $T_{\nu}^{ \ 
\alpha\beta} T_{\alpha\mu\beta}$ are also not-symmetric, however they will give 
the same contributions with some of the terms we have listed above and 
therefore we ignore them. Moreover, terms of the form $\nabla_\mu \phi 
\nabla_\nu \phi, T_\mu T_\nu,T_{\alpha\beta\mu} T^{\alpha\beta}_{ \ \ \nu}$ and 
$ T^{\alpha \ \beta}_{ \ \mu} T_{\beta \nu \alpha}$ are symmetric and therefore 
we are not allowed to use them.
Finally, terms of the form $T_{\beta\mu\nu} 
\nabla^\beta \phi, T_{\beta\mu\nu} T^\beta$ and $\nabla^\beta T_{\beta\mu\nu}$ 
are anti-symmetric. Building on the work described in Ref.~\cite{Bahamonde:2019shr,Bahamonde:2019ipm}, these terms can be understood to produce second order terms in the equations of motion, and to be part of a generalized Horndeski theory in which torsion has replaced curvature as the geometric mediator of gravitation. While further work is needed to fully assemble the foundational interpretation of these terms, they are already well-placed in the TG literature.

Now, we are in position to apply the field redefinitions (\ref{eq:L1} - \ref{eq:L7}) to the leading torsional action, whose transformation is given by (\ref{eq:redef11a}). As an example, we consider the field redefinition in (\ref{eq:L2}). The computation is long-winded but straightforward. After several integration by parts, (\ref{eq:redef11a}) becomes
\begin{align}
    \tilde{e} \tilde{T} & =  e \bigg\{  T  + \frac{1}{8} \alpha_1(\phi) \lo^\alpha \phi \Big[  2 T_{\alpha }{}^{\beta \kappa \
} \qty(-4 \, T_{\beta \kappa }{}^{\lambda } + T^{\lambda }{}_{\beta \kappa \
}) T_{\lambda} + T_{\alpha} 
\qty(T_{\kappa \lambda \mu } T^{\kappa \lambda \mu } + 2 \, T^{\kappa 
\lambda \mu } T_{\lambda \kappa \mu } - 12 \, T^{\lambda} T_{\lambda} + 8 \, \nabla_{\lambda 
}T^{\lambda }) \nonumber
\\&  +4 \qty(- T_{\beta \kappa
}{}^{\lambda } T^{\beta }{}_{\alpha }{}^{\kappa } T_{\lambda} - T^{\beta }{}_{\alpha }{}^{\kappa } T_{\kappa \beta 
}{}^{\lambda } T_{\lambda} + T^{\kappa } \qty(- 2 \, \nabla_{\alpha }T_{\kappa }  - 
\nabla_{\lambda }T_{\alpha \kappa }{}^{\lambda } - \nabla_{\lambda 
}T_{\kappa \alpha }{}^{\lambda } + \nabla_{\lambda }T^{\lambda 
}{}_{\alpha \kappa })) \Big] \bigg\}\,.\label{eq:redef11aa}
\end{align} 
It is now clear that the effect of the field redefinition is to generate operators at next to leading order in the action. Repeated field redefinitions in terms of the operators in (\ref{eq:L1} - \ref{eq:L7}) can help us identify which higher-order derivative terms can be removed from the EFT, as we will see in Sec. \ref{sec:build_eft}.

\subsection{Field redefinitions of the scalar field} \label{sec:scalarredf}
  
 We close this section presenting the field redefinitions of the scalar field. 
These are significantly easier,  and  are expressed as
\begin{equation}
    \tilde{\phi} \rightarrow \phi + \frac{1}{\Lambda^2} b(\phi) \Phi\,,\label{eq:phir1}
\end{equation}
where $\Phi$ denotes a scalar object  containing at most two derivatives. We can 
use the following anzatses for $\Phi$
\begin{align}
    &  \Phi = b_1(\phi) \, \lc{\square} \phi\,,\label{eq:Lp1}\\
    & \Phi = b_2(\phi) \, \nabla^\alpha T_\alpha\,,\label{eq:Lp2}\\
    & \Phi = b_3(\phi) \, T_\alpha \lo^\alpha \phi\,,\label{eq:Lp3}\\
    & \Phi = b_4(\phi) \, T_\alpha^{ \ \beta \delta}  T^\alpha_{ \ \beta\delta}\,,\label{eq:Lp4}\\
    & \Phi = b_5(\phi) \, T_\alpha^{ \ \beta \delta}  T_{\beta \ \delta}^{ \ \alpha}\,,\label{eq:Lp5}\\
    & \Phi = b_6(\phi) \, T_\alpha T^\alpha\,.\label{eq:Lp6}
\end{align}
We mention that one  needs to  track the different derivatives carefully, since 
 the mixing of teleparallel and Levi-Civita derivatives will be taken into 
account in the integration by parts.

\section{The effective field theory of TEGR}
 \label{sec:build_eft}
 
 In this section we will construct the   EFT for Teleparallel gravity with a 
 minimally-coupled  scalar field around a general background, which is the main 
focus of the 
present work, applying the formalism of Weinberg's seminal work \cite{2008}. 
Additionally, for the first time we attempt to examine the similarities and 
differences between curvature and torsional gravity, at next-to-leading order in 
the perturbative expansion.   

At first glance, one may expect the two theories to be identical, since at a 
fundamental level the degrees of freedom and symmetries of gravity should not 
change. Indeed, for cosmological purposes the  construction of the EFT, for 
both theories, involves a scalar field and the graviton, which must obey 
Lorentz invariance. While this may be the case, there are important structural 
differences between the two theories when building an EFT, from a bottom-up 
approach, which must be carefully addressed. 

In order to construct the EFT we need to write down all the 
terms at next-to-leading-order containing the torsion tensor, derivatives of the 
torsion tensor and derivatives of the scalar field, accompanied by a generic 
coupling, which is a function of the scalar field. Here, we can have, at most, 
two derivatives acting on the tetrad and the scalar field at any given time. 
Finally, these operators are suppressed by a heavy mass scale $\Lambda$ 
(for instance $\Lambda \sim m_P$ for an inflationary set-up).
Lastly, note that in principle there  should also be parity-violating 
contributions too, which will not be considered in this work and will  be 
examined separately.

For example, in the case of curvature gravity with a minimally-coupled scalar 
field, one needs to write   the most general Lorentz-invariant Lagrangian 
that is a function of the metric, the Riemann tensor and its contractions, and 
the scalar fields and its derivatives, namely
\begin{equation}
\begin{split}
    \Delta L_{\text{GR},\phi} & \big(g_{\mu\nu},\phi, f(\phi), \pd_\alpha \phi, 
\lc{\square} \phi, \lc{R}, \lc{R}_{\mu\nu}, \lc{R}^\mu_{ \ \nu \rho \sigma} + 
\cdots \big)\,,
\end{split}
\end{equation}
where the Riemann tensor is expressed in terms of the Levi-Civita connection.   
On the other hand,  the torsional EFT with a single scalar field will have a 
different structure, namely
\begin{equation}
     \Delta L_{\text{TG},\phi}   \big(g_{\mu\nu},\phi, f(\phi), \lo_\alpha 
\phi, 
\lc{\square} \phi\,,   
     e^A_{ \ \mu}, T_\mu, T^\mu_{ \ \nu\rho}, \nabla_\alpha T_\mu,  
\nabla_\alpha T^\mu_{ \ \nu\rho} + \cdots \big)\,.
 \end{equation}
The first difference between the two EFTs is that in the latter we need to use two 
connections, that is both the Levi-Civita and the teleparallel 
one. The second difference is that the two EFTs are comprised by objects which 
have different symmetries. For example, the metric is symmetric under exchange 
of two indices while the tetrad is an asymmetric object. Additionally, the 
torsion tensor is anti-symmetric under exchange of the last two indices, while 
the Ricci tensor is symmetric under exchange of its indices, etc. 
The third important difference is that curvature tensors are comprised by two 
derivatives acting on the metric, while the torsion tensor has only one 
derivative acting on the tetrad. All these features imply that we 
are allowed to write   significantly more terms in the torsional EFT 
comparing to those of curvature gravity, which was already known from 
\cite{Li:2018ixg}.

We start with operators constructed purely from derivatives of the scalar 
field, which have been  considered in  \cite{2008}. These are
\begin{equation}
    \Delta L_\phi = f_1(\phi) \qty(g^{\mu\nu} \accentset{\circ}{\nabla}_\mu \phi 
\lc{\nabla}_\nu \phi)^2 + f_2(\phi) g^{\rho\sigma} \lo_\rho \phi \lo_\sigma \phi 
\lc{\square} \phi + f_3(\phi) (\lc{\square} \phi)^2\,,
\label{eq:phiop}
\end{equation}
where it was shown that the last two terms amount to the field redefinition 
of the scalar field \eqref{eq:Lp1} and therefore only the first term 
contributes to the EFT. We proceed to the more complicated operators, which involve torsion terms. 
In particular, due to the single-derivative nature of torsion tensor, the 
torsion-scalar EFT is cumbersome, and thus for convenience we split it 
into several Lagrangians.  
In total, we consider the following Lagrangian  
\begin{equation}
    \Delta L_{TG}=\Delta L_1 +\Delta L_2 + \Delta L_3 +\Delta L_4  +\Delta L_5 
+\Delta L_6 +\Delta L_7 +\Delta L_8 +\Delta L_9+\Delta 
L_{10}\,.\label{eq:lagtot1}
\end{equation}
The Lagrangian $\Delta L_1$ consists of operators with one torsion tensor and 
one scalar field, and with at most two derivatives acting on the tetrad and the 
scalar field, that is 
\begin{align}
    \Delta L_1 & = g_{1}(\phi) \,   \nabla_{\kappa }T^{\kappa } \lc{\square}\phi 
+ g_{2}(\phi) \,  \nabla_{\kappa }T_{\alpha \beta }{}^{\kappa } 
\lc{\nabla}^{\beta } \lc{\nabla}^{\alpha }\phi + g_3(\phi) \,  \nabla_{\kappa 
}T_{\beta \alpha }{}^{\kappa } \lo^{\beta }\lo^{\alpha }\phi + g_4(\phi) \,  
\nabla^{\kappa }T_{\kappa \alpha }{}^{\beta} \lo_{\beta }\lo^{\alpha 
}\phi\nonumber\\
    & + g_{5}(\phi) \, \nabla_{\alpha }T_{\beta } \lo^{\beta }\lo^{\alpha }\phi 
+  g_{6}(\phi) \, \nabla_{\beta }T_{\alpha} \lo^{\beta }\lo^{\alpha 
}\phi\,.\label{eq:lag1}
\end{align}
For convenience, from now on we will refer to operators in terms of their 
coupling.  The operator $g_1$ can be removed by the field redefinition in 
(\ref{eq:Lp2}), while the operators $g_2 - g_6$ cannot be removed by a field 
redefinition. The Lagrangian $\Delta L_2$ contains operators with two torsion tensors and at 
most two derivatives acting on the tetrad. It is given by
\begin{align}
    \Delta L_2 & = g_{7}(\phi) \, \nabla^{\kappa }T^{\beta }  \nabla_{\lambda 
}T^{\lambda }{}_{\beta \kappa } + g_{8}(\phi) \, \nabla^{\kappa }T^{\beta } 
\nabla_{\lambda }T_{\kappa \beta }{}^{\lambda } + g_{9}(\phi) \, \nabla^{\kappa 
}T^{\beta } \nabla_{\lambda }T_{\beta \kappa }{}^{\lambda } + g_{10}(\phi) \, 
\nabla_{\beta }T^{\beta } \nabla_{\lambda }T^{\lambda } + g_{11}(\phi) \, 
\nabla_{\beta }T_{\kappa } \nabla^{\kappa }T^{\beta } \nonumber\\
    & + g_{12}(\phi) \,\nabla_{\alpha }T^{\alpha \beta \kappa } \nabla_{\lambda 
}T^{\lambda }{}_{\beta \kappa } + g_{13}(\phi) \, \nabla_\alpha T^\lambda 
\nabla^\alpha T_\lambda + g_{14}(\phi) \, \nabla_{\alpha }T^{\beta  \kappa 
\alpha } \nabla_{\lambda }T_{\beta \kappa }{}^{\lambda } + g_{15}(\phi) \, 
\nabla_{\alpha }T^{\kappa  \beta \alpha} \nabla_{\lambda }T_{\beta \kappa 
}{}^{\lambda } \nonumber\\
    & + g_{16}(\phi) \, \nabla_{\alpha }T^{\alpha \beta \kappa} \nabla_{\lambda 
}T_{\beta \kappa }{}^{\lambda }  + g_{17}(\phi) \, \nabla_{\alpha }T^{\beta 
\kappa \lambda } \nabla_{\lambda }T_{ \ \beta \kappa }^{\alpha } +  g_{18}(\phi) 
\, \nabla_{\alpha }T^{\lambda \beta \kappa } \nabla_{\lambda }T_{ \ \beta \kappa 
}^{\alpha } + g_{19}(\phi) \, \nabla_{\alpha }T^{\beta \kappa \lambda } 
\nabla_{\lambda }T_{\beta \kappa }{}^{\alpha } \nonumber\\
    &+  g_{20}(\phi) \, \nabla_{\alpha }T^{\kappa \beta \lambda }  
\nabla_{\lambda }T_{\beta \kappa }{}^{\alpha }  + g_{21}(\phi) \, \nabla_\mu 
T_{\nu\rho\sigma} \, \nabla^\mu T^{\nu \rho\sigma} + g_{22}(\phi) \, \nabla_\mu 
T_{\nu\rho\sigma} \, \nabla^\mu T^{\rho\nu\sigma} \nonumber\\
    & =  g_{13}(\phi) \, \nabla_\alpha T^\lambda \nabla^\alpha T_\lambda + 
g_{14}(\phi) \, \nabla_{\alpha }T^{\beta \kappa \alpha } \nabla_{\lambda 
}T_{\beta \kappa }{}^{\lambda } + g_{15}(\phi) \, \nabla_{\alpha }T^{\kappa  
\beta \alpha} \nabla_{\lambda }T_{\beta \kappa }{}^{\lambda }  + g_{16}(\phi) \, 
\nabla_{\alpha }T^{\alpha \beta \kappa} \nabla_{\lambda }T_{\beta \kappa 
}{}^{\lambda }\,. \label{eq:lag2}
\end{align}
 The $g_{12}$ term can be removed by the field redefinition in (\ref{eq:L5}). 
The rest operators cannot be removed by a field redefinition. The Lagrangian $\Delta L_3$   contains terms with two torsion tensors 
coupled to a term with two derivatives acting on the scalar field, and it is 
given by
\begin{align}
    \Delta L_3 & =   g_{23}(\phi) \, T^{\beta } T^{\lambda } \nabla_{\lambda 
}\lo_{\beta }\phi + g_{24}(\phi) \, T^\lambda T_\lambda \lc{\square} \phi + 
g_{25}(\phi) \, T_{\alpha \beta }{}^{\kappa } T_{\kappa} \lo^{\beta }\lo^{\alpha 
}\phi + g_{26}(\phi) \, T_{\beta \alpha }{}^{\kappa } T_{\kappa } \lo^{\beta 
}\lo^{\alpha }\phi \nonumber\\
    & + g_{27}(\phi) \, T^{\kappa  }{}_{\alpha \beta } T_{\kappa } \lo^{\beta 
}\lo^{\alpha }\phi + g_{28}(\phi) \, T^{\alpha \beta \kappa } T_{\beta \alpha 
}{}^{\lambda } \lo_{\lambda }\lo_{\kappa }\phi + g_{29}(\phi) \, T^{\alpha \beta 
\kappa } T_{\beta \kappa }{}^{\lambda } \lo_{\lambda }\lo_{\alpha }\phi 
\nonumber\\
    & + g_{30}(\phi) \, T^{\alpha \beta \kappa } T^{\lambda }{}_{\beta \kappa }  
\lo_{\lambda }\lo_{\alpha }\phi + g_{31}(\phi) \, T_{\lambda }{}^{\beta \kappa } 
T_{\beta \kappa \alpha } \lo^{\lambda }\lo^{\alpha }\phi + g_{32}(\phi) \,  
T^{\alpha \beta \kappa } T_{\alpha \beta }{}^{\lambda } \lo_{\lambda 
}\lo_{\kappa }\phi \nonumber\\
    & + g_{33}(\phi) \, T_{\alpha \beta \kappa } T^{\alpha \beta \kappa }  
\lc{\square}\phi + g_{34}(\phi) \,  T^{\alpha \beta \kappa } T_{\beta \alpha 
\kappa } \lc{\square}\phi\,. \label{eq:lag3}
\end{align}
Operators $g_{23},g_{24},g_{28},g_{29},g_{30},g_{32},g_{33},g_{34}$ can be 
removed by the field redefinition in (\ref{eq:L6}). Operator $g_{33}$ can also 
be removed by the field redefinition in (\ref{eq:Lp4}),   operator $g_{34}$ 
can   be removed using (\ref{eq:Lp5}), while   
$g_{24}$ can   be removed using (\ref{eq:Lp6}). The Lagrangian $\Delta L_4$ contains operators with two torsion terms and one 
scalar field, and with at most two derivatives acting on the tetrad and one on 
the scalar field. It is given by
\begin{equation}\begin{split}
\Delta L_4 & =   g_{35}(\phi) \, T^\lambda \nabla_\alpha T_\lambda \lo^\alpha \phi  
+  g_{36}(\phi) \, T^{\lambda }  \nabla_{\lambda }T_{\alpha } \lo^{\alpha }\phi
+ g_{37}(\phi) \, T_{\alpha }  \nabla_{\lambda }T^{\lambda } \lo^{\alpha }\phi  
\\& + g_{38}(\phi)  \, T_{\alpha }{}^{\beta \kappa }  \nabla_{\kappa }T_{\beta} 
\lo^{\alpha }\phi 
+ g_{39}(\phi) \, T^{\beta }{}_{\alpha }{}^{\kappa }   \nabla_{\beta }T_{\kappa 
} \lo^{\alpha }\phi 
+ g_{40}(\phi) \, T^{\beta }{}_{\alpha }{}^{\kappa }  \nabla_{\kappa }T_{\beta}  
\lo^{\alpha }\phi 
\\& + g_{41}(\phi) \, T^{\kappa }   \nabla_{\lambda }T_{\alpha \kappa 
}{}^{\lambda } \lo^{\alpha }\phi 
+ g_{42}(\phi) \,  T^{\kappa }  \nabla_{\lambda }T_{\kappa \alpha }{}^{\lambda } 
 \lo^{\alpha }\phi 
+  g_{43}(\phi) \, T^{\kappa }  \nabla_{\lambda }T^{\lambda }{}_{\alpha \kappa}  
\lo^{\alpha }\phi 
\\& + g_{44}(\phi)  \, T^{\beta \kappa \lambda }  \nabla_{\beta }T_{\kappa 
\alpha \lambda } \lo^{\alpha }\phi
+ g_{45}(\phi) \, T^{\beta \kappa \lambda}  \nabla_{\beta }T_{\alpha \kappa 
\lambda }  \lo^{\alpha }\phi 
+ g_{46}(\phi) \, T^{\beta \kappa \lambda }  \nabla_{\lambda }T_{\kappa \alpha 
\beta }  \lo^{\alpha }\phi
\\& + g_{47}(\phi) \, T^{\beta \kappa \lambda }  \nabla_{\lambda }T_{\alpha 
\beta  \kappa }  \lo^{\alpha }\phi 
+ g_{48}(\phi) \, T_{\beta \kappa }{}^{\lambda }  \nabla_{\lambda } T^{\beta 
}{}_{\alpha }{}^{\kappa }   \lo^{\alpha }\phi 
+ g_{49}(\phi) \, T_{\alpha }{}^{\beta \kappa }    \nabla_{\lambda }T_{\beta 
\kappa }{}^{\lambda } \lo^{\alpha }\phi 
\\& + g_{50}(\phi) \, T^{\beta }{}_{\alpha }{}^{\kappa }   \nabla_{\lambda 
}T_{\kappa \beta }{}^{\lambda } \lo^{\alpha }\phi  
+ g_{51}(\phi) \, T^{\beta }{}_{\alpha }{}^{\kappa }   \nabla_{\lambda }T_{\beta 
\kappa }{}^{\lambda } \lo^{\alpha }\phi 
+ g_{52}(\phi) \, T_{\alpha }{}^{\beta \kappa }   \nabla_{\lambda}T^{\lambda 
}{}_{\beta \kappa } \lo^{\alpha }\phi 
\\& + g_{53}(\phi) \, T^{\beta }{}_{\alpha }{}^{\kappa }   \nabla_{\lambda 
}T^{\lambda }{}_{\beta \kappa } \lo^{\alpha }\phi 
+ g_{54}(\phi) \, T^{\beta \kappa\lambda } \nabla_{\alpha }T_{\beta 
\kappa\lambda }   \lo^{\alpha }\phi 
+ g_{55}(\phi) \,  T^{\beta \kappa\lambda } \nabla_{\alpha }T_{ \kappa \beta 
\lambda } \lo^{\alpha }\phi\,.
\label{eq:lag4}
\end{split}\end{equation}
Operators $g_{36},g_{44},g_{46},g_{48}$ can be removed by the field redefinition 
in (\ref{eq:L2}), operators $g_{37}, g_{41},g_{43}$ can be removed using 
(\ref{eq:L1}), while operator $g_{42}$ can be removed by the field 
redefinitions in (\ref{eq:L1}) and (\ref{eq:L2}). Next, we have terms with  two torsion tensors and two scalar fields derivatively 
coupled to the torsion tensors. These terms cannot be removed by field 
redefinitions. In particular, we have
\begin{equation}\begin{split}
\Delta L_5 & =  g_{56}(\phi) \, T^{\delta }  T_{\delta } \lo_{\alpha }\phi 
\lo^{\alpha }\phi
+ g_{57}(\phi) \, T_{\alpha} T_{\beta}  \lo^{\alpha }\phi \lo^{\beta }\phi
+ g_{58}(\phi) \,  T_{\alpha \beta }{}^{\delta } T_{\delta} \lo^{\alpha }\phi 
\lo^{\beta }\phi
+ g_{59}(\phi) \, T_{\beta \delta \epsilon } T^{\beta \delta \epsilon }  
\lo_{\alpha }\phi \lo^{\alpha }\phi
\\&+ g_{60}(\phi) \, T^{\beta \delta \epsilon } T_{\delta \beta \epsilon }  
\lo_{\alpha }\phi \lo^{\alpha }\phi
+ g_{61}(\phi) \,  T_{\alpha }{}^{\delta \epsilon } T_{\beta \delta \epsilon }  
\lo^{\alpha }\phi \lo^{\beta }\phi
+ g_{62}(\phi) \, T_{\alpha }{}^{\delta \epsilon } T_{\delta \beta \epsilon }  
\lo^{\alpha }\phi \lo^{\beta }\phi
\\& + g_{63}(\phi) \, T_{\delta \beta \epsilon } T^{\delta }{}_{\alpha 
}{}^{\epsilon }  \lo^{\alpha }\phi \lo^{\beta }\phi
+ g_{64}(\phi) \, T^{\delta }{}_{\alpha }{}^{\epsilon } T_{\epsilon \beta \delta 
}  \lo^{\alpha }\phi \lo^{\beta }\phi\,. 
\label{eq:lag5}
\end{split}\end{equation}
Similarly,  we have terms with three torsion tensors. Since these terms are  
related, one to another, by integration by parts, one should write them 
carefully. Thus we have 
\begin{equation}\begin{split}
\Delta L_6 & =  g_{65}(\phi) \,T^{\lambda } T_{\lambda }  \nabla_{\alpha 
}T^{\alpha } +  g_{66}(\phi) \, T^{\beta } T^{\kappa }{}_{\beta }{}^{\lambda } 
\nabla_{\lambda }T_{\kappa } +g_{67}(\phi) \, T^{\alpha \beta \kappa } T_{\beta 
\kappa }{}^{\lambda } \nabla_{\lambda }T_{\alpha} +g_{68}(\phi) \, T^{\lambda } 
T^{\beta }{}_{\lambda }{}^{\kappa } \nabla_{\alpha }T^{\alpha }{}_{\beta\kappa }
\\&+ g_{69}(\phi) \, T^{\alpha \beta \kappa } T^{\lambda }{}_{\beta \kappa } 
\nabla_{\lambda }T_{\alpha} + g_{70}(\phi) \, T^{\kappa } T^{\alpha }{}_{\kappa 
}{}^{\beta} \nabla_{\lambda }T_{\alpha \beta }{}^{\lambda } +g_{71}(\phi) \, 
T^{\beta } T^{\kappa }{}_{\beta }{}^{\lambda } \nabla_{\mu }T_{\lambda \kappa 
}{}^{\mu } +g_{72}(\phi) \, T_{\alpha \beta \kappa } T^{\alpha \beta \kappa } 
\nabla_{\mu}T^{\mu } \\& +g_{73}(\phi) \, T^{\alpha \beta \kappa } T_{\beta 
\alpha \kappa } \nabla_{\mu }T^{\mu } +  g_{74}(\phi) \, T^{\alpha \beta \delta 
} T_{\beta \delta }{}^{\epsilon } \nabla_{\zeta }T^{\zeta }{}_{\alpha \epsilon } 
+g_{75}(\phi) \,T^{\alpha \beta \delta } T_{\beta }{}^{\epsilon \zeta }  
\nabla_{\delta }T_{\alpha \epsilon \zeta } +g_{76}(\phi) \, T^{\alpha \beta 
\delta } T_{\beta \delta }{}^{\epsilon } \nabla_{\zeta }T_{\alpha \epsilon 
}{}^{\zeta }
\\& +  g_{77}(\phi) \, T^{\alpha \beta \delta } T_{\beta \delta }{}^{\epsilon }  
\nabla_{\zeta }T_{\epsilon \alpha }{}^{\zeta } +  g_{78}(\phi) \, T^{\alpha 
\beta \delta } T_{\beta }{}^{\epsilon \zeta } \nabla_{\zeta }T_{\alpha \delta 
\epsilon } +  g_{79}(\phi) \,T^{\alpha \beta \delta } T_{\beta }{}^{\epsilon 
\zeta } \nabla_{\zeta }T_{\delta \alpha \epsilon } + g_{80}(\phi) \, T^{\beta } 
T^{\kappa } \nabla_{\lambda }T_{\beta \kappa }{}^{\lambda } 
\\& +  g_{81}(\phi) \, T^{\alpha \beta \delta } T_{\beta }{}^{\epsilon \zeta }  
\nabla_{\alpha }T_{\delta \epsilon \zeta } +  g_{82}(\phi) \,T^{\alpha \beta 
\delta } T^{\epsilon }{}_{\beta }{}^{\zeta } \nabla_{\epsilon }T_{\alpha \delta 
\zeta } 
+ g_{83}(\phi) \, T^{\alpha \beta \delta } T^{\epsilon }{}_{\beta }{}^{\zeta }  
\nabla_{\epsilon }T_{\delta \alpha \zeta }\,.
\label{eq:lag6}
\end{split}\end{equation}
Operator $g_{65}$ can be removed by the field redefinition in (\ref{eq:L3}), 
operators $g_{66},g_{68},g_{70},g_{71}$ can be removed using (\ref{eq:L3}),
 $g_{67},g_{73}$ can be removed using
(\ref{eq:L7}),   $g_{74}$ can be removed using (\ref{eq:L5}) and 
(\ref{eq:L7}), and   $g_{76}, g_{77}$ can be removed  using (\ref{eq:L7}). 
Again, due to how the operators above are related to others by integration by 
parts, we find that some field redefinitions re-introduce certain terms. For 
this reason we decide  not to include these field redefinitions, which is just 
our choice to  represent the theory. Additionally, we have operators with three torsion tensors and one scalar 
field derivatively coupled to torsion, given by
\begin{equation}\begin{split}
\Delta L_7 & =  g_{84}(\phi) \, T_\lambda T^\lambda  T_\alpha \lo^\alpha \phi 
+ g_{85}(\phi) \, T^{\beta }{}_{\alpha }{}^{\kappa } T_{\beta } T_{\kappa } \lo^{\alpha }\phi 
+ g_{86}(\phi) \, T_{\beta \kappa }{}^{\lambda }  T_{\alpha }{}^{\beta \kappa }  
T_{\lambda} \lo^{\alpha }\phi 
+ g_{87}(\phi) \, T_{\beta \kappa }{}^{\lambda } T^{\beta }{}_{\alpha }{}^{\kappa } T_{\lambda} \lo^{\alpha }\phi 
\\& + g_{88}(\phi) \, T_{\kappa \beta }{}^{\lambda } T^{\beta }{}_{\alpha }{}^{\kappa }   T_{\lambda} \lo^{\alpha }\phi 
+ g_{89}(\phi) \, T^{\lambda }{}_{\beta \kappa } T_{\alpha }{}^{\beta \kappa }  T_{\lambda } \lo^{\alpha }\phi  
+ g_{90}(\phi) \, T^{\beta }{}_{\alpha }{}^{\kappa} T^{\lambda }{}_{\beta \kappa} T_{\lambda} \lo^{\alpha }\phi
\\& + g_{91}(\phi) \,  T^{\kappa \lambda \mu } T_{\lambda \kappa \mu } T_{\alpha } \lo^{\alpha }\phi 
+ g_{92}(\phi) \, T_{\kappa \lambda \mu }  T^{\kappa \lambda \mu } T_\alpha 
\lo^{\alpha }\phi 
+ g_{93}(\phi) \, T^{\beta }{}_{\alpha }{}^{\kappa } T_{\kappa \lambda \mu } T_{\beta }{}^{\lambda \mu }   \lo^{\alpha }\phi 
\\&  + g_{94}(\phi) \, T^{\beta }{}_{\alpha }{}^{\kappa }  T_{\kappa 
}{}^{\lambda \mu } T_{\lambda \beta \mu } \lo^{\alpha }\phi 
 + g_{95}(\phi) \, T^{\beta }{}_{\alpha }{}^{\kappa } T_{\lambda \kappa \mu } T^{\lambda }{}_{\beta }{}^{\mu } \lo^{\alpha }\phi 
 + g_{96}(\phi) \, T^{\beta }{}_{\alpha }{}^{\kappa }  T_{\mu \kappa \lambda } 
T^{\lambda }{}_{\beta }{}^{\mu } \lo^{\alpha }\phi  
\\& + g_{97}(\phi) \,T_{\beta }{}^{\lambda \mu }  T^{\beta }{}_{\alpha 
}{}^{\kappa } T_{\lambda \kappa \mu } \lo^{\alpha }\phi
+ g_{98}(\phi) \, T_{\alpha }{}^{\beta \kappa }  T_{\beta }{}^{\lambda \mu } 
T_{\lambda \kappa\mu } \lo^{\alpha }\phi\,.
\label{eq:lag8}
\end{split}\end{equation}
Operators $g_{84}, g_{91},  g_{92} $ can be removed by the field 
redefinitions in (\ref{eq:L1}) and (\ref{eq:L2}), $g_{85},g_{93}, 
g_{94}, g_{95}, g_{96}$ can be removed using
(\ref{eq:L2}), and $g_{86}, g_{87}, g_{88}, g_{89}$ can be removed using 
(\ref{eq:L1}). We proceed to operators with four torsion tensors. We separate them 
into two Lagrangians. The first is given by
\begin{equation}\begin{split}
\Delta L_8 & =  g_{99}(\phi) \, T^{\beta } T_{\beta } T^{\mu } T_{\mu } 
+ g_{100}(\phi) \, T^{\beta } T_{\beta } T_{\lambda \mu \nu } T^{\lambda \mu \nu }  
+ g_{101}(\phi) \, T^{\beta } T_{\beta} T^{\lambda \mu \nu } T_{\mu \lambda \nu } 
+ g_{102}(\phi) \, T^{\beta } T_{\zeta } T_{\beta }{}^{\delta \epsilon } T_{\delta \epsilon }{}^{\zeta } 
\\& + g_{103}(\phi) \, T^{\beta } T_{\zeta } T_{\delta \epsilon }{}^{\zeta } T^{\delta }{}_{\beta }{}^{\epsilon } 
+ g_{104}(\phi) \, T^{\beta } T_{\zeta } T^{\delta }{}_{\beta }{}^{\epsilon } T_{\epsilon \delta }{}^{\zeta } 
+ g_{105}(\phi) \, T^{\beta } T_{\zeta} T_{\beta }{}^{\delta \epsilon } T^{\zeta }{}_{\delta \epsilon } 
+ g_{106}(\phi) \, T^{\beta }  T^{\kappa }{}_{\beta }{}^{\lambda } T_{\lambda \mu \nu }T_{\kappa }{}^{\mu \nu } 
\\& + g_{107}(\phi) \, T^{\beta } T^{\kappa }{}_{\beta }{}^{\lambda } T_{\lambda }{}^{\mu \nu } T_{\mu \kappa \nu } 
+ g_{108}(\phi) \, T^{\beta } T^{\kappa }{}_{\beta }{}^{\lambda } T_{\mu \lambda \nu } T^{\mu }{}_{\kappa }{}^{\nu } 
+ g_{109}(\phi) \, T^{\beta } T^{\kappa }{}_{\beta }{}^{\lambda }  T_{\nu \lambda \mu } T^{\mu }{}_{\kappa }{}^{\nu } 
\\&+g_{110}(\phi) \, T^{\beta } T_{\beta }{}^{\delta \epsilon } T_{\delta }{}^{\zeta \eta } T_{\zeta \epsilon \eta } +g_{111}(\phi) \, T^{\beta } T_{\delta }{}^{\zeta \eta } T^{\delta }{}_{\beta }{}^{\epsilon } T_{\zeta \epsilon \eta }\,.
\label{eq:lag9}
\end{split}\end{equation}
Operators $g_{99},g_{100},g_{106},g_{107},g_{108},g_{109}$ can be removed by 
the field redefinitions in (\ref{eq:L3}),  $g_{101},$ can be removed using 
(\ref{eq:L3}) and (\ref{eq:L7}), and   $g_{102}$ can be removed using  
(\ref{eq:L7}). The second Lagrangian containing four torsion tensors is given 
by
\begin{equation}\begin{split}
\Delta L_{9} & =  g_{112}(\phi) \, T_{\alpha }{}^{\epsilon \zeta }
 T^{\alpha \beta \delta } T_{\beta \epsilon }{}^{\eta } T_{\delta \zeta \eta }
+ g_{113}(\phi) \,  T_{\alpha }{}^{\epsilon \zeta } T^{\alpha \beta \delta } 
T_{\beta \delta }{}^{\eta } T_{\epsilon \zeta \eta }
+ g_{114}(\phi) \,  T_{\alpha \beta }{}^{\epsilon } T^{\alpha \beta \delta } 
T_{\delta }{}^{\zeta \eta } T_{\epsilon \zeta \eta }
\\& + g_{115}(\phi) \, T^{\alpha \beta \delta } T_{\beta \alpha }{}^{\epsilon } T_{\delta }{}^{\zeta \eta } T_{\epsilon \zeta \eta }
+ g_{116}(\phi) \, T_{\alpha \beta \delta } T^{\alpha \beta \delta } T_{\epsilon \zeta \eta } T^{\epsilon \zeta \eta }
+ g_{117}(\phi) \,  T^{\alpha \beta \delta } T_{\beta }{}^{\epsilon \zeta } 
T_{\delta \epsilon }{}^{\eta } T_{\zeta \alpha \eta }
\\&+ g_{118}(\phi)  \, T^{\alpha \beta \delta } T_{\beta \delta }{}^{\epsilon } 
T_{\epsilon }{}^{\zeta \eta } T_{\zeta \alpha \eta } 
+ g_{119}(\phi) \,  T_{\alpha }{}^{\epsilon \zeta } T^{\alpha \beta \delta } 
T_{\beta \epsilon }{}^{\eta } T_{\zeta \delta \eta }
+ g_{120}(\phi)  \, T_{\alpha \beta }{}^{\epsilon } T^{\alpha \beta \delta } 
T_{\delta }{}^{\zeta \eta } T_{\zeta \epsilon \eta } 
\\& + g_{121}(\phi) \,  T^{\alpha \beta \delta } T_{\beta \alpha }{}^{\epsilon } T_{\delta }{}^{\zeta \eta } T_{\zeta \epsilon \eta } 
+  g_{122}(\phi)  \, T_{\alpha \beta \delta } T^{\alpha \beta \delta } 
T^{\epsilon \zeta \eta } T_{\zeta \epsilon \eta } 
+ g_{123}(\phi) \,  T^{\alpha \beta \delta } T_{\beta \alpha \delta } 
T^{\epsilon \zeta \eta } T_{\zeta \epsilon \eta } 
\\& +  g_{124}(\phi) \, T_{\alpha \beta }{}^{\epsilon } T^{\alpha \beta \delta } T_{\zeta \epsilon \eta } T^{\zeta }{}_{\delta }{}^{\eta }
+ g_{125}(\phi) \,  T^{\alpha \beta \delta } T_{\beta }{}^{\epsilon \zeta } 
T_{\epsilon \delta }{}^{\eta } T_{\eta \alpha \zeta }
+ g_{126}(\phi) \, T_{\alpha }{}^{\epsilon \zeta } T^{\alpha \beta \delta } T_{\beta \epsilon }{}^{\eta } T_{\eta \delta \zeta }
\\& + g_{127}(\phi)  \, T_{\alpha }{}^{\epsilon \zeta } T^{\alpha \beta \delta } 
T_{\beta \delta }{}^{\eta } T_{\eta \epsilon \zeta } 
+ g_{128}(\phi)  \, T_{\alpha \beta }{}^{\epsilon } T^{\alpha \beta \delta } 
T^{\zeta }{}_{\delta }{}^{\eta } T_{\eta \epsilon \zeta }
+ g_{129}(\phi)  \, T^{\alpha \beta \delta } T_{\beta \alpha }{}^{\epsilon } 
T^{\zeta }{}_{\delta }{}^{\eta } T_{\eta \epsilon \zeta }
\\& + g_{130}(\phi) \, T_{\alpha }{}^{\epsilon \zeta } T^{\alpha \beta \delta } T_{\eta \epsilon \zeta } T^{\eta }{}_{\beta \delta }
+ g_{131}(\phi) \,  T_{\alpha }{}^{\epsilon \zeta } T^{\alpha \beta \delta } 
T_{\eta \delta \zeta } T^{\eta }{}_{\beta \epsilon }\,,
\label{eq:lag10}
\end{split}\end{equation}
where operators $g_{118},g_{120},g_{121},g_{122},g_{123},g_{127}$ can be 
removed using (\ref{eq:L7}). Finally, the Lagrangian $\Delta L_{10}$ contains operators 
with one torsion tensor and two scalar fields derivatively coupled to torsion, 
namely
\begin{equation}\begin{split}
\Delta L_{10} &  = g_{132}(\phi) \,  T_{\beta} \lo_{\alpha }\lo^{\beta }\phi 
\lo^{\alpha }\phi
+ g_{133}(\phi) \,  T_{\alpha} \nabla^{\alpha }\phi \square \phi
+ g_{134}(\phi) \,  T_{\beta } \lo^{\alpha }\phi \lo^{\beta }\lo_{\alpha }\phi 
+ g_{135}(\phi) \,  T_{\alpha \beta \delta } \lo^{\alpha }\phi \lo^{\delta 
}\lo^{\beta }\phi
\\& + g_{136}(\phi)  \, T_{\beta \alpha \delta } \lo^{\alpha }\phi \lo^{\delta 
}\lo^{\beta }\phi
+ g_{137}(\phi) \,  T_{\delta \alpha \beta } \lo^{\alpha }\phi \lo^{\delta 
}\lo^{\beta }\phi\,.
\label{eq:lag11}
\end{split}\end{equation}
These operators cannot be removed by a field redefinition apart from $g_{133}$ 
which can be removed using (\ref{eq:Lp3}). Inserting all the above into (\ref{eq:lagtot1}), and removing all 
operators that can be removed, we obtain
\begin{equation}\begin{split}
\Delta L_{TG} & = 
 g_{2}(\phi) \,  \nabla_{\kappa }T_{\alpha \beta }{}^{\kappa } 
\lc{\nabla}^{\beta } \lc{\nabla}^{\alpha }\phi
+ g_3(\phi) \,  \nabla_{\kappa }T_{\beta \alpha }{}^{\kappa } \lo^{\beta 
}\lo^{\alpha }\phi
+ g_4(\phi) \,  \nabla^{\kappa }T_{\kappa \alpha }{}^{\beta} \lo^{\beta 
}\lo_{\alpha }\phi
\\& + g_{5}(\phi) \, \nabla_{\alpha }T_{\beta } \lo^{\beta }\lo^{\alpha }\phi
+ g_{6}(\phi) \, \nabla_{\beta }T_{\alpha} \lo^{\beta }\lo^{\alpha }\phi + 
g_{7}(\phi) \, \nabla^{\kappa }T^{\beta } \nabla_{\lambda }T^{\lambda }{}_{\beta 
\kappa } 
+ g_{8}(\phi) \, \nabla^{\kappa }T^{\beta } \nabla_{\lambda }T_{\kappa \beta 
}{}^{\lambda } 
\\& + g_{9}(\phi) \, \nabla^{\kappa }T^{\beta } \nabla_{\lambda }T_{\beta \kappa 
}{}^{\lambda } 
+ g_{10}(\phi) \, \nabla_{\beta }T^{\beta } \nabla_{\lambda }T^{\lambda } 
+ g_{11}(\phi) \, \nabla_{\beta }T_{\kappa } \nabla^{\kappa }T^{\beta } 
\\&  + g_{13}(\phi) \, \nabla_\alpha T^\lambda \nabla^\alpha T_\lambda 
+ g_{14}(\phi) \, \nabla_{\alpha }T^{\beta \kappa \alpha } \nabla_{\lambda 
}T_{\beta \kappa }{}^{\lambda } 
+ g_{15}(\phi) \, \nabla_{\alpha }T^{\kappa  \beta \alpha} \nabla_{\lambda 
}T_{\beta \kappa }{}^{\lambda } 
\\& + g_{16}(\phi) \, \nabla_{\alpha }T^{\alpha \beta \kappa} \nabla_{\lambda 
}T_{\beta \kappa }{}^{\lambda } 
 + g_{17}(\phi) \, \nabla_{\alpha }T^{\beta \kappa \lambda } \nabla_{\lambda 
}T_{ \ \beta \kappa }^{\alpha } 
+  g_{18}(\phi) \, \nabla_{\alpha }T^{\lambda \beta \kappa } \nabla_{\lambda 
}T_{ \ \beta \kappa }^{\alpha } 
+ g_{19}(\phi) \, \nabla_{\alpha }T^{\beta \kappa \lambda } \nabla_{\lambda 
}T_{\beta \kappa }{}^{\alpha } 
\\& +  g_{20}(\phi) \, \nabla_{\alpha }T^{\kappa \beta \lambda }  
\nabla_{\lambda }T_{\beta \kappa }{}^{\alpha } 
+ g_{21}(\phi) \, \nabla_\mu T_{\nu\rho\sigma} \, \nabla^\mu T^{\nu \rho\sigma} 
+ g_{22}(\phi) \, \nabla_\mu T_{\nu\rho\sigma} \, \nabla^\mu T^{\rho\nu\sigma} + 
g_{25}(\phi) \, T_{\alpha \beta }{}^{\kappa } T_{\kappa} \lo^{\beta }\lo^{\alpha 
}\phi 
\\& + g_{26}(\phi) \, T_{\beta \alpha }{}^{\kappa } T_{\kappa } \lo^{\beta 
}\lo^{\alpha }\phi + g_{27}(\phi) \, T^{\kappa }{}_{\alpha \beta } T_{\kappa } 
\lo^{\beta }\lo^{\alpha }\phi + g_{31}(\phi) \, T_{\lambda }{}^{\beta \kappa } 
T_{\beta \kappa \alpha } \lo^{\lambda }\lo^{\alpha }\phi + g_{38}(\phi) \, 
T_{\alpha }{}^{\beta \kappa }  \nabla_{\kappa }T_{\beta} \lo^{\alpha }\phi 
\\&+ g_{39}(\phi) \, T^{\beta }{}_{\alpha }{}^{\kappa }  \nabla_{\beta 
}T_{\kappa } \lo^{\alpha }\phi 
+ g_{40}(\phi) \, T^{\beta }{}_{\alpha }{}^{\kappa }  \nabla_{\kappa }T_{\beta} 
\lo^{\alpha }\phi 
+ g_{45}(\phi) \, T^{\beta \kappa \lambda}  \nabla_{\beta }T_{\alpha \kappa 
\lambda } \lo^{\alpha }\phi 
+ g_{47}(\phi) \, T^{\beta \kappa \lambda }  \nabla_{\lambda }T_{\alpha \beta  
\kappa } \lo^{\alpha }\phi 
\\&+ g_{49}(\phi) \, T_{\alpha }{}^{\beta \kappa }   \nabla_{\lambda }T_{\beta 
\kappa }{}^{\lambda } \lo^{\alpha }\phi 
+ g_{50}(\phi) \, T^{\beta }{}_{\alpha }{}^{\kappa }  \nabla_{\lambda }T_{\kappa 
\beta }{}^{\lambda } \lo^{\alpha }\phi  
+ g_{51}(\phi) \, T^{\beta }{}_{\alpha }{}^{\kappa }  \nabla_{\lambda }T_{\beta 
\kappa }{}^{\lambda } \lo^{\alpha }\phi 
\\& + g_{52}(\phi) \, T_{\alpha }{}^{\beta \kappa }  \nabla_{\lambda}T^{\lambda 
}{}_{\beta \kappa } \lo^{\alpha }\phi 
+ g_{53}(\phi) \, T^{\beta }{}_{\alpha }{}^{\kappa }  \nabla_{\lambda 
}T^{\lambda }{}_{\beta \kappa } \lo^{\alpha }\phi 
+ g_{54}(\phi) \, T^{\beta \kappa\lambda } \nabla_{\alpha }T_{\beta 
\kappa\lambda }  \lo^{\alpha }\phi 
\\& + g_{55}(\phi) \, T^{\beta \kappa\lambda } \nabla_{\alpha }T_{ \kappa \beta 
\lambda } \lo^{\alpha }\phi  + g_{56}(\phi) \, T^{\delta } T_{\delta } 
\lo_{\alpha }\phi \lo^{\alpha }\phi
+ g_{57}(\phi) \, T_{\alpha} T_{\beta} \lo^{\alpha }\phi \lo^{\beta }\phi
+ g_{58}(\phi) \, T_{\alpha \beta }{}^{\delta } T_{\delta} \lo^{\alpha }\phi 
\lo^{\beta }\phi
\\& + g_{59}(\phi) \, T_{\beta \delta \epsilon } T^{\beta \delta \epsilon } 
\lo_{\alpha }\phi \lo^{\alpha }\phi
+ g_{60}(\phi) \, T^{\beta \delta \epsilon } T_{\delta \beta \epsilon } 
\lo_{\alpha }\phi \lo^{\alpha }\phi
+ g_{61}(\phi) \,  T_{\alpha }{}^{\delta \epsilon } T_{\beta \delta \epsilon } 
\lo^{\alpha }\phi \lo^{\beta }\phi
\\&+ g_{62}(\phi) \, T_{\alpha }{}^{\delta \epsilon } T_{\delta \beta \epsilon } 
\lo^{\alpha }\phi \lo^{\beta }\phi
 + g_{63}(\phi) \, T_{\delta \beta \epsilon } T^{\delta }{}_{\alpha 
}{}^{\epsilon } \lo^{\alpha }\phi \lo^{\beta }\phi
+ g_{64}(\phi) \, T^{\delta }{}_{\alpha }{}^{\epsilon } T_{\epsilon \beta \delta 
}
\lo^{\alpha }\phi \lo^{\beta }\phi 
\\& + g_{69}(\phi) \, T^{\alpha \beta \kappa } T^{\lambda }{}_{\beta \kappa } 
\nabla_{\lambda }T_{\alpha} +g_{72}(\phi) \, T_{\alpha \beta \kappa } T^{\alpha 
\beta \kappa } \nabla_{\mu}T^{\mu }
\\& +g_{75}(\phi) \,T^{\alpha \beta \delta } T_{\beta }{}^{\epsilon \zeta } 
\nabla_{\delta }T_{\alpha \epsilon \zeta } +  g_{78}(\phi) \, T^{\alpha \beta 
\delta } T_{\beta }{}^{\epsilon \zeta } \nabla_{\zeta }T_{\alpha \delta \epsilon 
} +  g_{79}(\phi) \,T^{\alpha \beta \delta } T_{\beta }{}^{\epsilon \zeta } 
\nabla_{\zeta }T_{\delta \alpha \epsilon } + g_{80}(\phi) \, T^{\beta } 
T^{\kappa } \nabla_{\lambda }T_{\beta \kappa }{}^{\lambda } 
\\& +  g_{81}(\phi) \, T^{\alpha \beta \delta } T_{\beta }{}^{\epsilon \zeta } 
\nabla_{\alpha }T_{\delta \epsilon \zeta } +  g_{82}(\phi) \,T^{\alpha \beta 
\delta } T^{\epsilon }{}_{\beta }{}^{\zeta } \nabla_{\epsilon }T_{\alpha \delta 
\zeta } 
+ g_{83}(\phi) \, T^{\alpha \beta \delta } T^{\epsilon }{}_{\beta }{}^{\zeta } 
\nabla_{\epsilon }T_{\delta \alpha \zeta } +  g_{90}(\phi) \, T^{\beta 
}{}_{\alpha }{}^{\kappa} T^{\lambda }{}_{\beta \kappa} T_{\lambda} \lo^\alpha 
\phi
\\& + g_{97}(\phi) \,T_{\beta }{}^{\lambda \mu } T^{\beta }{}_{\alpha 
}{}^{\kappa } T_{\lambda \kappa \mu } \lo^{\alpha }\phi
+ g_{98}(\phi) \, T_{\alpha }{}^{\beta \kappa } T_{\beta }{}^{\lambda \mu } 
T_{\lambda \kappa\mu }   \lo^{\alpha }\phi +  g_{103}(\phi) \, T^{\beta } 
T_{\zeta } T_{\delta \epsilon }{}^{\zeta } T^{\delta }{}_{\beta }{}^{\epsilon } 
\\& + g_{104}(\phi) \, T^{\beta } T_{\zeta } T^{\delta }{}_{\beta }{}^{\epsilon 
} T_{\epsilon \delta }{}^{\zeta } 
+ g_{105}(\phi) \, T^{\beta } T_{\zeta} T_{\beta }{}^{\delta \epsilon } T^{\zeta 
}{}_{\delta \epsilon } +g_{110}(\phi) \, T^{\beta } T_{\beta }{}^{\delta 
\epsilon } T_{\delta }{}^{\zeta \eta } T_{\zeta \epsilon \eta }
 +g_{111}(\phi) \, T^{\beta } T_{\delta }{}^{\zeta \eta } T^{\delta }{}_{\beta 
}{}^{\epsilon } T_{\zeta \epsilon \eta } 
 \\&+ g_{112}(\phi) \, T_{\alpha }{}^{\epsilon \zeta } T^{\alpha \beta \delta } 
T_{\beta \epsilon }{}^{\eta } T_{\delta \zeta \eta }
+ g_{113}(\phi) \, T_{\alpha }{}^{\epsilon \zeta } T^{\alpha \beta \delta } 
T_{\beta \delta }{}^{\eta } T_{\epsilon \zeta \eta }
+ g_{114}(\phi) \, T_{\alpha \beta }{}^{\epsilon } T^{\alpha \beta \delta } 
T_{\delta }{}^{\zeta \eta } T_{\epsilon \zeta \eta }
\\& + g_{115}(\phi) \, T^{\alpha \beta \delta } T_{\beta \alpha }{}^{\epsilon } 
T_{\delta }{}^{\zeta \eta } T_{\epsilon \zeta \eta }
+ g_{116}(\phi) \, T_{\alpha \beta \delta } T^{\alpha \beta \delta } T_{\epsilon 
\zeta \eta } T^{\epsilon \zeta \eta }
+ g_{117}(\phi) \, T^{\alpha \beta \delta } T_{\beta }{}^{\epsilon \zeta } 
T_{\delta \epsilon }{}^{\eta } T_{\zeta \alpha \eta }
\\& + g_{119}(\phi) \, T_{\alpha }{}^{\epsilon \zeta } T^{\alpha \beta \delta } 
T_{\beta \epsilon }{}^{\eta } T_{\zeta \delta \eta } +  g_{124}(\phi) \, 
T_{\alpha \beta }{}^{\epsilon } T^{\alpha \beta \delta } T_{\zeta \epsilon \eta 
} T^{\zeta }{}_{\delta }{}^{\eta }
+ g_{125}(\phi) \, T^{\alpha \beta \delta } T_{\beta }{}^{\epsilon \zeta } 
T_{\epsilon \delta }{}^{\eta } T_{\eta \alpha \zeta }
\\& + g_{126}(\phi) \, T_{\alpha }{}^{\epsilon \zeta } T^{\alpha \beta \delta } 
T_{\beta \epsilon }{}^{\eta } T_{\eta \delta \zeta }  
+ g_{128}(\phi) \, T_{\alpha \beta }{}^{\epsilon } T^{\alpha \beta \delta } 
T^{\zeta }{}_{\delta }{}^{\eta } T_{\eta \epsilon \zeta }
 + g_{129}(\phi) \, T^{\alpha \beta \delta } T_{\beta \alpha }{}^{\epsilon } 
T^{\zeta }{}_{\delta }{}^{\eta } T_{\eta \epsilon \zeta }
\\& + g_{130}(\phi) \, T_{\alpha }{}^{\epsilon \zeta } T^{\alpha \beta \delta } 
T_{\eta \epsilon \zeta } T^{\eta }{}_{\beta \delta }
+ g_{131}(\phi) \, T_{\alpha }{}^{\epsilon \zeta } T^{\alpha \beta \delta } 
T_{\eta \delta \zeta } T^{\eta }{}_{\beta \epsilon } + g_{132}(\phi) \,  
T_{\beta} \nabla_{\alpha }\lo^{\beta }\phi \lo^{\alpha }\phi
\\& + g_{134}(\phi) \, T_{\beta } \lo^{\alpha }\phi \lo^{\beta }\lo_{\alpha 
}\phi 
+ g_{135}(\phi) \, T_{\alpha \beta \delta } \lo^{\alpha }\phi \lo^{\delta 
}\lo^{\beta }\phi
 + g_{136}(\phi) \, T_{\beta \alpha \delta } \lo^{\alpha }\phi \lo^{\delta 
}\lo^{\beta }\phi
\\& + g_{137}(\phi) \, T_{\delta \alpha \beta } \lo^{\alpha }\phi \lo^{\delta 
}\lo^{\beta }\phi\,. 
\label{eq:lagTG}
\end{split}\end{equation}
In summary, the EFT action at next-to-leading order is written as
\begin{equation}
    S = -m_P^2 \int \dd^4{x} \left\{e T+ e \frac{1}{\Lambda^2}\Delta L_{TG} + 
\sqrt{-g} \left[\lo_\mu \phi \lo^\mu \phi - V(\phi)\right] + \sqrt{-g} 
\frac{1}{\Lambda^2} \Delta L_\phi\right\}\,,\label{eq:EFT}
\end{equation}
with
\begin{equation}
    \Delta L_\phi = f_1(\phi) \qty(g^{\mu\nu} \accentset{\circ}{\nabla}_\mu \phi 
\lc{\nabla}_\nu \phi)^2\,,\label{eq:lagtot1a}
\end{equation}
and where    we have now made explicit the mass scale $\Lambda^2$  and 
treat the coefficients as dimensional couplings of $\mathcal{O}(1)$. 

We have managed to construct the general EFT of TEGR. As we mentioned above, 
the Lagrangian in (\ref{eq:lagTG}) will lead to higher than second-order 
equations of motion. These can be dealt by additional field redefinitions, on a 
case-by-case basis, by fixing the background geometry. Hence, in the next 
section we consider a cosmological   Friedmann-Lema\^{i}tre-Robertson-Walker 
spacetime.

\section{Application to Cosmology} \label{sec:cosmology}

In the previous section we constructed the  EFT framework for TEGR with a 
single scalar field. The higher-order operators are suppressed by the mass 
scale $\Lambda$, which is the energy scale where new physics may appear. 
Nevertheless, as we mentioned, remaining at the general background does not 
allow     to remove all problematic operators, since the usual 
methods for ghost reduction in curvature gravity do not straightforwardly apply 
to TEGR. Actually, this was expected,  since the 
regular approach is designed primarily to handle curvature-based terms, while 
teleparallel gravity produces novel terms that require us to take an ad hoc 
approach to removing such ghosts. However, one can handle these  operators 
on a case-by-case basis by considering   a specific 
background spacetime.

In this section,  we focus on the flat Friedmann–Lema\^{i}tre–Robertson–Walker 
(FLRW) geometry, which is used for cosmological applications. In particular, we 
consider the metric 
\begin{equation}\label{eq:FLRW_metric}
    ds^2 = N(t)^2dt^2 - a(t)^2(dx^2+dy^2+dz^2)\,,
\end{equation}
where $N$ is the lapse function and $a(t)$ is the scale factor. This metric can 
arise from  the tetrad
\begin{equation}
    h^a{}_{\mu}=\textrm{diag}(N(t),a(t),a(t),a(t))\,,
\end{equation}
which satisfies the Weitzenb\"{o}ck gauge \cite{Hohmann:2020zre,Krssak:2018ywd}.
Note that in this case the torsion scalar takes the simple form 
\cite{Cai:2015emx}
\begin{equation}
    T = -6H^2\,,
\end{equation}
where $H=\dot{a}/a$ is the Hubble function and dots denote derivatives with 
respect to $t$.

In our approach  we consider the point-like Lagrangian shown in 
 \eqref{eq:EFT}. Given that the metric  \eqref{eq:FLRW_metric} is maximally 
symmetric, we can use the Euler-Lagrange equations to extract the equations of 
motion using the variables $\{N(t), a(t), \phi(t)\}$. In order to make the first 
form of the new Friedmann equations comprehensible, we equate all the couplings. 
The result of the respective variations are the Friedmann equations and the 
Klein-Gordon equation, namely
\begin{align}
    & H^2 + \frac{g(\phi)}{\Lambda^2}  A_1 +  \frac{g'(\phi)}{\Lambda^2}  B_1 
=\frac{1}{6} \dot{\phi} ^2 + f_1(\phi) \frac{ \dot{\phi}^4}{ \Lambda ^2}+ 
U_1\,, \label{eq:fr1a}\\
    & \dot{H}+\frac{3 H^2}{2} + \frac{g(\phi)}{\Lambda^2}   C_1 + 
\frac{g'(\phi)}{\Lambda^2}  D_1=-\frac{\dot{\phi} ^2}{4} - f_1(\phi) 
\frac{\dot{\phi} ^4}{ 2\Lambda ^2} +U_2\,,\label{eq:fr5a}\\
    &\ddot{\phi} +3 H \dot{\phi}  + \frac{3 \dot{\phi}^2}{\Lambda ^2}  
\qty[f_1'(\phi) \dot{\phi}^2+4 f_1(\phi) H \dot{\phi}+4 f_1(\phi) \ddot{\phi}] 
+ \frac{g(\phi)}{\Lambda^2}  E_1 + \frac{g'(\phi)}{\Lambda^2}  F_1 + U_3 = 0\,, 
\label{eq:fr9a}
\end{align}
with primes denoting derivatives with respect to $\phi$ and
where the detailed forms of the various coefficient functions are
 given in   Appendix \ref{Appendix}.
These are  the governing equations at the background level, arising from the EFT of TEGR with a  minimally-coupled scalar field.

Let us now come to the question of the Introduction, and examine whether the 
above EFT theory is equivalent with the standard EFT of General Relativity with 
a minimally-coupled scalar field. In the latter case, the action, up to 
next-to-leading order, is \cite{2008}
\begin{equation}
   S = m_P^2 \int \dd^4{x} \sqrt{-g}\qty[ \frac{1}{2} \lc{R} +  
\frac{f_0(\phi)}{\Lambda^2} \lc{R}_{\mu\nu\rho\sigma} \lc{R}^{\mu\nu\rho\sigma} 
+  \qty(\frac{1}{2}\lc{\nabla}_\mu \phi \lc{\nabla}^\mu \phi - V(\phi)) +  
\frac{1}{\Lambda^2} f_1(\phi) \qty(g^{\mu\nu} \lc{\nabla}_\mu \phi 
\lc{\nabla}_\nu \phi)^2]\label{eq:EFT_GR},
\end{equation}
where covariant derivatives are calculated using the Levi-Civita connection 
(note that we choose to work with the Riemann tensor 
since the Weyl tensor vanishes in a conformally flat background). Note that for 
simplicity we do not consider the parity violating gravitational 
terms. Hence, the corresponding equations of motion are
\begin{align}
    & H^2 - \frac{f_0(\phi)}{\Lambda^2} A_2 -  \frac{f_0'(\phi)}{\Lambda^2}   
B_2 = \frac{1}{3} \frac{ \dot{\phi} ^2}{2}+\frac{f_1(\phi)  \dot{\phi} 
^4}{\Lambda ^2} + V_1\label{eq:R1a}\\
    & \dot{H}+\frac{3 }{2} H^2 -\frac{f_0(\phi)}{\Lambda^2} C_2 - 
\frac{f_0'(\phi)}{\Lambda^2} D_2 = -\frac{1}{2}\frac{ \dot{\phi} ^2}{2} - 
\frac{f_1(\phi)  \dot{\phi} ^4}{2 \Lambda ^2} +V_2 \label{eq:R4a} \\
    & \ddot{\phi}+ 
3 H \dot{\phi}  + \frac{3 \dot{\phi}^2}{\Lambda ^2}  
\qty[f_1'(\phi) \dot{\phi}^2+4 f_1(\phi) H \dot{\phi}+4 f_1(\phi) \ddot{\phi}]  
+\frac{f_0(\phi)}{\Lambda^2} E_2 + \frac{f_0'(\phi)}{\Lambda^2} F_2+ V_3 = 0, 
\label{eq:R8a}
\end{align}
where the detailed forms of the coefficient functions are presented in Appendix \ref{Appendix}. Given the breadth of the new terms produced by the EFT of TEGR, a diverse plethora of models can be constructed, many of which one would expect to produce many of the characteristics of current cosmology such as late time accelerated expansion. These models may also offer some resolution to the current contention on the Hubble tension \cite{Abdalla:2022yfr}. Additionally, the the EFT of torsion gravity, constructed here, can provide richer phenomenology for inflation, being a highly energetic theory and it could be especially important in regimes where the strength of gravity is strong, such as blackholes.

Let us now compare the two cases. As we know, GR with a minimally coupled 
scalar field and TEGR with  a minimally coupled 
scalar field   lead to identical equations (we stress that this does not happen 
when non-minimal couplings are considered, since scalar-tensor and 
scalar-torsion theories are different \cite{Geng:2011aj}, but this is not the 
case of the present manuscript).
Furthermore, the leading order contributions in both GR 
and TEGR are the same for their corresponding EFTs. 

However, the higher order 
structure    does produce some differences, since $A_1 \neq A_2, 
B_1 \neq B_2, \cdots F_1, \neq F_2$ (see Appendix \ref{Appendix}) and this is 
the main result of the present work.   Nevertheless, we mention that although 
different,  
 it may still be difficult to experimentally distinguish one from the other, 
since  they are highly suppressed by the mass scale $\Lambda$, placing these 
contributions at the sub-percent level. For instance, in  inflationary 
considerations one has $\Lambda \sim M_P$ where $M_P ~ 2.4 \times 10^{18} $ 
GeV. Assuming $\mathcal{O}(1)$ coefficients then the higher-order contributions 
of the GR and TEGR EFTs with a minimally-coupled scalar field will be $\ll 1$ 
and therefore it will be difficult to be experimentally distinguishable.

\section{Conclusions}
\label{sec:conclusion}

In this work we constructed the Effective Field Theory of the teleparallel 
equivalent of general relativity, which is an equivalent description of 
gravity using torsion. Although the EFT for curvature-based gravity is well 
studied, this is not the case for torsional theories, since this requires a
significant extension of the regular approach to effective field theories.
One of our  motivations was to examine whether the degrees of freedom that get 
excited at high energies in the corresponding EFTs, behave differently, thus 
breaking the known equivalence of the two theories at the classical, low-energy 
limit.

After presenting the basics of TEGR, we proceeded to necessary field 
redefinitions of  the scalar field, as well as of the tetrads, since contrary 
to curvature-based gravity, where metric redefinitions are not needed, in 
torsional gravity the tetrad and torsion redefinitions are necessary since 
 the equations of motion depend on both the Lorentz and general coordinates of the manifold structure. We continued by writing down all 
the terms at next-to-leading-order, containing the torsion tensor, derivatives 
of the torsion tensor and derivatives of the scalar field, accompanied by a 
generic coupling, which is a function of the scalar field, where all operators   
are suppressed by a  scale $\Lambda$. Although using the field redefinitions we 
were able to remove many terms, the resulting EFT has significantly more terms 
comparing to the EFT of curvature gravity, since in contrast to the symmetric 
metric the tetrad has no symmetries and moreover the torsion tensor contains one 
tetrad derivative in contrast to the two metric derivatives of the curvature 
tensor. Finally, after  constructing the general EFT of TEGR, as an example we 
applied it in a cosmological framework, which allowed us to handle and remove  
possible remaining higher-derivative terms, resulting to second-order equations of motion.

Interestingly enough, although GR and TEGR with a minimally-coupled scalar 
field are completely equivalent at the level of classical equations, we found 
that this is not the case in their corresponding EFTs, since they possess some 
differences, minor but non-zero. Hence, we do verify that at higher energies 
the excitation and the features of the extra degrees of freedom are slightly 
different in the two theories, thus making them theoretically distinguishable.
Nevertheless, we mention that these differences are suppressed by  the mass 
scale $\Lambda$, and thus it is not guaranteed that they could be measured in 
future experiments and observations.  

It would be interesting to investigate  in detail the high-energy inflationary 
realization in the two cases, in which we expect the EFT structure to affect 
the observables, and examine whether we can acquire distinguishable features. Indeed the exact form of the dependencies on the terms in the EFT of TEGR (shown in Appendix \ref{Appendix}) may reveal more significant distinguishing features through their respective predictions on this and other early Universe physics dynamics.
Additionally, we should proceed to the detailed perturbation analysis and see 
whether we obtain different structures  in some phenomenological regime. 
Finally, one should proceed to the detailed consideration of  parity violating 
terms in the EFT of TEGR. These interesting and necessary investigations lie 
beyond the scope of this work, and are left for future projects.

\begin{acknowledgments}
The authors would like to acknowledge the contribution of the COST Action 
CA21136 ``Addressing observational tensions in cosmology with systematics and 
fundamental physics (CosmoVerse)''. M.M. is supported in part by the National
Research Foundation of Korea Grant 2019R1A2C2085023.

\end{acknowledgments}

\appendix

\section {   The functions appearing in 
Eqs.~(\ref{eq:fr1a})-(\ref{eq:fr9a}) and (\ref{eq:R1a})-(\ref{eq:R8a})} 
\label{Appendix}

In this Appendix we present the exact form of the functions that appear in 
Eqs.~(\ref{eq:fr1a})-(\ref{eq:fr9a}) and (\ref{eq:R1a})-(\ref{eq:R8a}). We 
have: 
\begin{align}
 U_1 & = \frac{V}{3} -\frac{1}{4 \Lambda ^2} \Big\{2 V' \left[\dot{\phi} 
\left(4 H g'(\phi)+50 g(\phi) H\right)+30 g(\phi) H^2+g(\phi) 
 \dot{\phi}  ^2\right]+124 H V g'(\phi) \dot{\phi}-4 V g'(\phi) 
 \dot{\phi}^2
   \nonumber\\&
   +972 g(\phi) H^2
   V+4 g(\phi) \dot{H} \left(2 V'+31 V\right)+8 g(\phi) H V'' \dot{\phi}-42 
g(\phi) H V \dot{\phi}-93 g(\phi) V^2-4
   g(\phi) V \ddot{\phi}
  \nonumber\\&
  +93 g(\phi) V  \dot{\phi}^2 \Big\}
\label{eq:fr2},
\\ 
A_1 & = \frac{1}{16 } \Big[-8940 H^4+8 \dot{\phi} \left(57 H^3-70 H 
\ddot{\phi}\right)+128 H^2 \ddot{\phi}+24
    \dot{\phi}^2 \left(\ddot{\phi}-123 H^2\right)
    \nonumber\\& -8 \dot{H} \left(594 H^2-32 H
   \dot{\phi}+35  \dot{\phi}^2\right)+84 H 
\dot{\phi}^3-93
    \dot{\phi}^4 \Big],
   \label{eq:fr3}
   \\
   B_1 & = \frac{\phi^\prime}{2 } \Big[-198 H^3+16 H^2 \dot{\phi}-35 H 
\dot{\phi}^2+\dot{\phi}^3  \Big],
\label{eq:fr4}
\end{align}

\begin{align}
U_2 & =  \frac{V}{2} - \frac{1}{8 \Lambda ^2} \Big\{4 V' \left\{2 H 
\left[3 
g'(\phi)+50 g(\phi)\right] \dot{\phi}-2 \left[g'(\phi)+2 g(\phi)\right] 
\dot{\phi}^2+g(\phi) \left(9 H^2-2 \ddot{\phi}\right)\right\}+400 H V 
g'(\phi) \dot{\phi}
   \nonumber\\& -10 V g'(\phi)
    \dot{\phi}^2+600 g(\phi) H^2 V+8 g(\phi) \dot{H} \left(3 V'+50 
V\right)+24 g(\phi) H V''
   \dot{\phi}-8 g(\phi) V''  \dot{\phi}^2+93 g(\phi) V^2
   \nonumber\\& -10 g(\phi) V \ddot{\phi}-93 g(\phi) V
    \dot{\phi}^2 \Big\}
   \label{eq:fr6},
      \\
   C_1 & =  \frac{1}{32 } 
   \Big[-12564 H^4+16 \dot{\phi} \left(9 H^3-176 H 
\ddot{\phi}\right)+72 H^2 \ddot{\phi}+\dot{\phi}^2 \left(60 
\ddot{\phi}-2112 H^2\right)
   \nonumber\\&
   -16 \dot{H} \left(1047 H^2-9 H \dot{\phi}+88 \dot{\phi}^2\right)+93 
 \dot{\phi}^4\Big]
   \label{eq:fr7},
   \\
   D_1 & =  \frac{\dot{\phi}}{8 } \Big[-1396 H^3+18 H^2 \dot{\phi}-352 H 
 \dot{\phi}^2+5 \dot{\phi}^3 \Big],
   \label{eq:fr8}
\end{align}

\begin{align}
U_3 & =  V -\frac{3}{4 \Lambda ^2} \Big\{ g'(\phi) \left[2 V' 
\left(\dot{\phi}^2-6 H^2\right)+V \left(-200 H^2+31
   V+31  \dot{\phi}^2\right)\right]
   \nonumber\\& +g(\phi) \Big[2 \left(-6 H^2 V''-15 H^2
   V+93 H V \dot{\phi}+V''  \dot{\phi}^2+31 V \ddot{\phi}\right)+V' 
\left(-224
   H^2+12 H \dot{\phi}+4 \ddot{\phi}+31  \dot{\phi}^2\right)
   \nonumber\\& -2 \dot{H} \left(4 V'+5
   V\right)\Big]\Big\}
    \label{eq:fr9},
    \\
    E_1 & =  -\frac{3}{4 } \Big[18 H^4-6 \dot{\phi} \left(176 H^3-5 H 
\ddot{\phi}\right)-352 H^2 \ddot{\phi}+\dot{\phi}^2 \left(45 H^2-93 
\ddot{\phi}\right)
   \nonumber\\& +\dot{H} \left(18 H^2-704 H \dot{\phi}
    +15
    \dot{\phi}^2\right)
    -93 H \dot{\phi}^3 \Big]
    \label{eq:fr10},
    \\
    F_1 & = - \frac{3}{16 } \qty[1396 H^4-704 H^2  \dot{\phi}^2+40 
H \dot{\phi}^3-93
    \dot{\phi}^4],
    \label{eq:fr11}
\end{align}

\begin{align}
V_1 & =V- \frac{1}{2 a^4 \Lambda ^2} \Big\{4 \left(a^4+1\right) H V f_0'(\phi) 
\dot{\phi}+f_0(\phi) V \left[\left(26
   a^4+2\right) H^2-\left(7 a^4+3\right) V+\left(7 a^4+3\right)
  \dot{\phi}^2\right]
   \nonumber\\& +4 \left(a^4+1\right) f_0(\phi) V \dot{H}+4
   \left(a^4+1\right) f_0(\phi) H V' \dot{\phi} \Big\}
\label{eq:R2},
\\
A_2 & = \frac{1}{8 a^4 } \Big[ 4 \left(19 a^4+23\right) H^4+4 \left(13 
a^4+1\right) H^2 \dot{\phi}^2+8 \left(a^4+1\right) \dot{H} \left(6 
H^2+\dot{\phi}^2\right)+16 \left(a^4+1\right) H \dot{\phi} 
\ddot{\phi}
   \nonumber\\& +\left(7
   a^4+3\right)  \dot{\phi}^4 \Big]
\label{eq:R3},
\\
B_2 & = \frac{1}{a^4 }\Big[\left(a^4+1\right) H \dot{\phi} \left(2 
H^2+\dot{\phi}^2\right)  \Big],
\label{eq:R4}
\end{align}

\begin{align}
V_2 & =  \frac{V}{2}  -\frac{1}{4 a^4 \Lambda ^2} \Big[4 \left(a^4+1\right) H V 
f_0'(\phi) \dot{\phi}+\left(3 a^4-1\right) f_0(\phi) V
   \left(2 H^2+V- \dot{\phi}^2\right)+4 \left(a^4+1\right) f_0(\phi)
   V \dot{H}
   \nonumber\\& +4 \left(a^4+1\right) f_0(\phi) H V' \dot{\phi}\Big]
\label{eq:R5},
\\
C_2 & = \frac{1}{16 a^4 } \Big\{36 \left(a^4-3\right) H^4+4 \left(3 
a^4-1\right) H^2  \dot{\phi}^2+8
   \dot{H} \left[6 \left(a^4+9\right) H^2 +\left(a^4+1\right)
  \dot{\phi}^2\right]
  \nonumber \\&  +16 \left(a^4+1\right) H \dot{\phi} \ddot{\phi}+\left(1-3
   a^4\right)  \dot{\phi}^4 \Big\}
\label{eq:R6},
\\
D_2 & =  \frac{1}{2 a^4} \Big[ 2 \left(a^4+9\right) H^3 
\dot{\phi}+\left(a^4+1\right) H \dot{\phi}^3 \Big],
\label{eq:R7}
\end{align}

\begin{align}
V_3 & =  V' + \frac{1}{2 a^4 \Lambda ^2} \Big\{-3 \left(a^4+1\right) V 
f_0'(\phi) \left(-2 H^2+V+\dot{\phi}^2\right)+3 \left(a^4+1\right) 
f_0(\phi) V' \left(2 H^2-2
   V- \dot{\phi}^2\right)
   \nonumber\\& -6 f_0(\phi) V \left[(3 a^4-1) H
   \dot{\phi}+(a^4+1) \ddot{\phi}\right]\Big\}
\label{eq:R9},
\\
E_2 & = \frac{3}{2 a^4 } \Big\{2 \left(a^4+1\right) H^2 \ddot{\phi}+\dot{\phi} 
\left[2 \left(3 a^4-1\right) H^3+4
   (a^4+1) H \dot{H}\right]+(3 a^4-1) H 
   \dot{\phi}^3
 +3 (a^4+1)  \dot{\phi}^2 \ddot{\phi} \Big\}
\label{eq:R10},
\\
F_2 & = \frac{3}{8 a^4 } \left[-4 \left(a^4+9\right) H^4+4 \left(a^4+1\right) 
H^2  \dot{\phi}^2+3
   \left(a^4+1\right)  \dot{\phi}^4\right],
\label{eq:R11}
\end{align}
where dots denote time derivatives and primes denote derivatives with respect 
to $\phi$.

\bibliographystyle{utphys}
\bibliography{refs}

\providecommand{\href}[2]{#2}\begingroup\raggedright\begin{thebibliography}{10}

\bibitem{misner1973gravitation}
C.~Misner, K.~Thorne, and J.~Wheeler, {\em Gravitation}.
\newblock No.~pt. 3 in Gravitation. W. H. Freeman, 1973.
\newblock \url{https://books.google.com.mt/books?id=w4Gigq3tY1kC}.

\bibitem{Clifton:2011jh}
T.~Clifton, P.~G. Ferreira, A.~Padilla, and C.~Skordis, ``{Modified Gravity and
  Cosmology},'' \href{http://dx.doi.org/10.1016/j.physrep.2012.01.001}{{\em
  Phys. Rept.} {\bf 513} (2012)  1--189},
\href{http://arxiv.org/abs/1106.2476}{{\tt arXiv:1106.2476 [astro-ph.CO]}}.

\bibitem{Aghanim:2018eyx}
{\bf Planck} Collaboration, N.~Aghanim {\em et al.}, ``{Planck 2018 results.
  VI. Cosmological parameters},''
\href{http://arxiv.org/abs/1807.06209}{{\tt arXiv:1807.06209 [astro-ph.CO]}}.

\bibitem{Riess:1998cb}
{\bf Supernova Search Team} Collaboration, A.~G. Riess {\em et al.},
  ``{Observational evidence from supernovae for an accelerating universe and a
  cosmological constant},'' \href{http://dx.doi.org/10.1086/300499}{{\em
  Astron.J.} {\bf 116} (1998)  1009--1038},
\href{http://arxiv.org/abs/astro-ph/9805201}{{\tt arXiv:astro-ph/9805201
  [astro-ph]}}.

\bibitem{Perlmutter:1998np}
{\bf Supernova Cosmology Project} Collaboration, S.~Perlmutter {\em et al.},
  ``{Measurements of Omega and Lambda from 42 high redshift supernovae},''
  \href{http://dx.doi.org/10.1086/307221}{{\em Astrophys.J.} {\bf 517} (1999)
  565--586},
\href{http://arxiv.org/abs/astro-ph/9812133}{{\tt arXiv:astro-ph/9812133
  [astro-ph]}}.

\bibitem{Abdalla:2022yfr}
E.~Abdalla {\em et al.}, ``{Cosmology intertwined: A review of the particle
  physics, astrophysics, and cosmology associated with the cosmological
  tensions and anomalies},''
  \href{http://dx.doi.org/10.1016/j.jheap.2022.04.002}{{\em JHEAp} {\bf 34}
  (2022)  49--211}, \href{http://arxiv.org/abs/2203.06142}{{\tt
  arXiv:2203.06142 [astro-ph.CO]}}.

\bibitem{Bernal:2016gxb}
J.~L. Bernal, L.~Verde, and A.~G. Riess, ``{The trouble with $H_0$},''
  \href{http://dx.doi.org/10.1088/1475-7516/2016/10/019}{{\em JCAP} {\bf 10}
  (2016)  019}, \href{http://arxiv.org/abs/1607.05617}{{\tt arXiv:1607.05617
  [astro-ph.CO]}}.

\bibitem{DiValentino:2020zio}
E.~Di~Valentino {\em et al.}, ``{Snowmass2021 - Letter of interest cosmology
  intertwined II: The hubble constant tension},''
  \href{http://dx.doi.org/10.1016/j.astropartphys.2021.102605}{{\em Astropart.
  Phys.} {\bf 131} (2021)  102605}, \href{http://arxiv.org/abs/2008.11284}{{\tt
  arXiv:2008.11284 [astro-ph.CO]}}.

\bibitem{DiValentino:2021izs}
E.~Di~Valentino, O.~Mena, S.~Pan, L.~Visinelli, W.~Yang, A.~Melchiorri, D.~F.
  Mota, A.~G. Riess, and J.~Silk, ``{In the realm of the Hubble
  tension\textemdash{}a review of solutions},''
  \href{http://dx.doi.org/10.1088/1361-6382/ac086d}{{\em Class. Quant. Grav.}
  {\bf 38} (2021) no.~15, 153001}, \href{http://arxiv.org/abs/2103.01183}{{\tt
  arXiv:2103.01183 [astro-ph.CO]}}.

\bibitem{DiValentino:2020vvd}
E.~Di~Valentino {\em et al.}, ``{Cosmology intertwined III: $f\sigma_8$ and
  $S_8$},'' \href{http://dx.doi.org/10.1016/j.astropartphys.2021.102604}{{\em
  Astropart. Phys.} {\bf 131} (2021)  102604},
  \href{http://arxiv.org/abs/2008.11285}{{\tt arXiv:2008.11285 [astro-ph.CO]}}.

\bibitem{DES:2017txv}
{\bf DES} Collaboration, T.~M.~C. Abbott {\em et al.}, ``{Dark Energy Survey
  Year 1 Results: A Precise H0 Estimate from DES Y1, BAO, and D/H Data},''
  \href{http://dx.doi.org/10.1093/mnras/sty1939}{{\em Mon. Not. Roy. Astron.
  Soc.} {\bf 480} (2018) no.~3, 3879--3888},
  \href{http://arxiv.org/abs/1711.00403}{{\tt arXiv:1711.00403 [astro-ph.CO]}}.

\bibitem{Riess:2021jrx}
A.~G. Riess {\em et al.}, ``{A Comprehensive Measurement of the Local Value of
  the Hubble Constant with 1 km s$^{-1}$ Mpc$^{-1}$ Uncertainty from the Hubble
  Space Telescope and the SH0ES Team},''
  \href{http://dx.doi.org/10.3847/2041-8213/ac5c5b}{{\em Astrophys. J. Lett.}
  {\bf 934} (2022) no.~1, L7}, \href{http://arxiv.org/abs/2112.04510}{{\tt
  arXiv:2112.04510 [astro-ph.CO]}}.

\bibitem{Brout:2021mpj}
D.~Brout {\em et al.}, ``{The Pantheon+ Analysis: SuperCal-Fragilistic Cross
  Calibration, Retrained SALT2 Light Curve Model, and Calibration Systematic
  Uncertainty},'' \href{http://arxiv.org/abs/2112.03864}{{\tt arXiv:2112.03864
  [astro-ph.CO]}}.

\bibitem{Scolnic:2021amr}
D.~Scolnic {\em et al.}, ``{The Pantheon+ Analysis: The Full Dataset and
  Light-Curve Release},'' \href{http://arxiv.org/abs/2112.03863}{{\tt
  arXiv:2112.03863 [astro-ph.CO]}}.

\bibitem{Baudis:2016qwx}
L.~Baudis, ``{Dark matter detection},''
\href{http://dx.doi.org/10.1088/0954-3899/43/4/044001}{{\em J. Phys.} {\bf G43}
  (2016) no.~4, 044001}.

\bibitem{Bertone:2004pz}
G.~Bertone, D.~Hooper, and J.~Silk, ``{Particle dark matter: Evidence,
  candidates and constraints},''
  \href{http://dx.doi.org/10.1016/j.physrep.2004.08.031}{{\em Phys. Rept.} {\bf
  405} (2005)  279--390},
\href{http://arxiv.org/abs/hep-ph/0404175}{{\tt arXiv:hep-ph/0404175
  [hep-ph]}}.

\bibitem{RevModPhys.61.1}
S.~Weinberg, ``{The Cosmological Constant Problem},''
  \href{http://dx.doi.org/10.1103/RevModPhys.61.1}{{\em Rev. Mod. Phys.} {\bf
  61} (1989)  1--23}.

\bibitem{Appleby:2018yci}
S.~Appleby and E.~V. Linder, ``{The Well-Tempered Cosmological Constant},''
  \href{http://dx.doi.org/10.1088/1475-7516/2018/07/034}{{\em JCAP} {\bf 07}
  (2018)  034}, \href{http://arxiv.org/abs/1805.00470}{{\tt arXiv:1805.00470
  [gr-qc]}}.

\bibitem{Ishak:2018his}
M.~Ishak, ``{Testing General Relativity in Cosmology},''
  \href{http://dx.doi.org/10.1007/s41114-018-0017-4}{{\em Living Rev. Rel.}
  {\bf 22} (2019) no.~1, 1}, \href{http://arxiv.org/abs/1806.10122}{{\tt
  arXiv:1806.10122 [astro-ph.CO]}}.

\bibitem{Addazi:2021xuf}
A.~Addazi {\em et al.}, ``{Quantum gravity phenomenology at the dawn of the
  multi-messenger era\textemdash{}A review},''
  \href{http://dx.doi.org/10.1016/j.ppnp.2022.103948}{{\em Prog. Part. Nucl.
  Phys.} {\bf 125} (2022)  103948}, \href{http://arxiv.org/abs/2111.05659}{{\tt
  arXiv:2111.05659 [hep-ph]}}.

\bibitem{CANTATA:2021ktz}
{\bf CANTATA} Collaboration, E.~N. Saridakis {\em et al.}, ``{Modified Gravity
  and Cosmology: An Update by the CANTATA Network},''
  \href{http://arxiv.org/abs/2105.12582}{{\tt arXiv:2105.12582 [gr-qc]}}.

\bibitem{Capozziello:2002rd}
S.~Capozziello, ``{Curvature quintessence},''
  \href{http://dx.doi.org/10.1142/S0218271802002025}{{\em Int. J. Mod. Phys. D}
  {\bf 11} (2002)  483--492}, \href{http://arxiv.org/abs/gr-qc/0201033}{{\tt
  arXiv:gr-qc/0201033}}.

\bibitem{Nicolis:2008in}
A.~Nicolis, R.~Rattazzi, and E.~Trincherini, ``{The Galileon as a local
  modification of gravity},''
  \href{http://dx.doi.org/10.1103/PhysRevD.79.064036}{{\em Phys. Rev. D} {\bf
  79} (2009)  064036}, \href{http://arxiv.org/abs/0811.2197}{{\tt
  arXiv:0811.2197 [hep-th]}}.

\bibitem{DeFelice:2010aj}
A.~De~Felice and S.~Tsujikawa, ``{f(R) theories},''
  \href{http://dx.doi.org/10.12942/lrr-2010-3}{{\em Living Rev. Rel.} {\bf 13}
  (2010)  3}, \href{http://arxiv.org/abs/1002.4928}{{\tt arXiv:1002.4928
  [gr-qc]}}.

\bibitem{Nojiri:2010wj}
S.~Nojiri and S.~D. Odintsov, ``{Unified cosmic history in modified gravity:
  from F(R) theory to Lorentz non-invariant models},''
  \href{http://dx.doi.org/10.1016/j.physrep.2011.04.001}{{\em Phys. Rept.} {\bf
  505} (2011)  59--144}, \href{http://arxiv.org/abs/1011.0544}{{\tt
  arXiv:1011.0544 [gr-qc]}}.

\bibitem{Mannheim:1988dj}
P.~D. Mannheim and D.~Kazanas, ``{Exact Vacuum Solution to Conformal Weyl
  Gravity and Galactic Rotation Curves},''
  \href{http://dx.doi.org/10.1086/167623}{{\em Astrophys. J.} {\bf 342} (1989)
  635--638}.

\bibitem{Horndeski:1974wa}
G.~W. Horndeski, ``{Second-order scalar-tensor field equations in a
  four-dimensional space},'' \href{http://dx.doi.org/10.1007/BF01807638}{{\em
  Int. J. Theor. Phys.} {\bf 10} (1974)  363--384}.

\bibitem{Wheeler:1985nh}
J.~T. Wheeler, ``{Symmetric Solutions to the Gauss-Bonnet Extended Einstein
  Equations},'' \href{http://dx.doi.org/10.1016/0550-3213(86)90268-3}{{\em
  Nucl. Phys. B} {\bf 268} (1986)  737--746}.

\bibitem{Nojiri:2005jg}
S.~Nojiri and S.~D. Odintsov, ``{Modified Gauss-Bonnet theory as gravitational
  alternative for dark energy},''
  \href{http://dx.doi.org/10.1016/j.physletb.2005.10.010}{{\em Phys. Lett. B}
  {\bf 631} (2005)  1--6}, \href{http://arxiv.org/abs/hep-th/0508049}{{\tt
  arXiv:hep-th/0508049}}.

\bibitem{DeFelice:2008wz}
A.~De~Felice and S.~Tsujikawa, ``{Construction of cosmologically viable f(G)
  dark energy models},''
  \href{http://dx.doi.org/10.1016/j.physletb.2009.03.060}{{\em Phys. Lett. B}
  {\bf 675} (2009)  1--8}, \href{http://arxiv.org/abs/0810.5712}{{\tt
  arXiv:0810.5712 [hep-th]}}.

\bibitem{lovelock1971einstein}
D.~Lovelock, ``The einstein tensor and its generalizations,'' {\em Journal of
  Mathematical Physics} {\bf 12} (1971) no.~3, 498--501.

\bibitem{Deruelle:1989fj}
N.~Deruelle and L.~Farina-Busto, ``{The Lovelock Gravitational Field Equations
  in Cosmology},'' \href{http://dx.doi.org/10.1103/PhysRevD.41.3696}{{\em Phys.
  Rev. D} {\bf 41} (1990)  3696}.

\bibitem{Unruh:1988in}
W.~G. Unruh, ``{A Unimodular Theory of Canonical Quantum Gravity},''
  \href{http://dx.doi.org/10.1103/PhysRevD.40.1048}{{\em Phys. Rev. D} {\bf 40}
  (1989)  1048}.

\bibitem{Harko:2011kv}
T.~Harko, F.~S.~N. Lobo, S.~Nojiri, and S.~D. Odintsov, ``{$f(R,T)$ gravity},''
  \href{http://dx.doi.org/10.1103/PhysRevD.84.024020}{{\em Phys. Rev. D} {\bf
  84} (2011)  024020}, \href{http://arxiv.org/abs/1104.2669}{{\tt
  arXiv:1104.2669 [gr-qc]}}.

\bibitem{Visser:2017gpz}
M.~Visser, ``{Rastall gravity is equivalent to Einstein gravity},''
  \href{http://dx.doi.org/10.1016/j.physletb.2018.05.028}{{\em Phys. Lett. B}
  {\bf 782} (2018)  83--86}, \href{http://arxiv.org/abs/1711.11500}{{\tt
  arXiv:1711.11500 [gr-qc]}}.

\bibitem{Capozziello:2010uv}
S.~Capozziello, J.~Matsumoto, S.~Nojiri, and S.~D. Odintsov, ``{Dark energy
  from modified gravity with Lagrange multipliers},''
  \href{http://dx.doi.org/10.1016/j.physletb.2010.08.030}{{\em Phys. Lett. B}
  {\bf 693} (2010)  198--208}, \href{http://arxiv.org/abs/1004.3691}{{\tt
  arXiv:1004.3691 [hep-th]}}.

\bibitem{Aldrovandi:2013wha}
R.~Aldrovandi and J.~G. Pereira,
  \href{http://dx.doi.org/10.1007/978-94-007-5143-9}{{\em {Teleparallel
  Gravity}}}, vol.~173.
\newblock Springer, Dordrecht,
2013.
\newblock

\bibitem{Cai:2015emx}
Y.-F. Cai, S.~Capozziello, M.~De~Laurentis, and E.~N. Saridakis, ``{$f(T)$
  teleparallel gravity and cosmology},''
  \href{http://dx.doi.org/10.1088/0034-4885/79/10/106901}{{\em Rept. Prog.
  Phys.} {\bf 79} (2016) no.~10, 106901},
\href{http://arxiv.org/abs/1511.07586}{{\tt arXiv:1511.07586 [gr-qc]}}.

\bibitem{Krssak:2018ywd}
M.~Krssak, R.~van~den Hoogen, J.~Pereira, C.~Böhmer, and A.~Coley,
  ``{Teleparallel theories of gravity: illuminating a fully invariant
  approach},'' \href{http://dx.doi.org/10.1088/1361-6382/ab2e1f}{{\em Class.
  Quant. Grav.} {\bf 36} (2019) no.~18, 183001},
  \href{http://arxiv.org/abs/1810.12932}{{\tt arXiv:1810.12932 [gr-qc]}}.

\bibitem{Bahamonde:2021gfp}
S.~Bahamonde, K.~F. Dialektopoulos, C.~Escamilla-Rivera, G.~Farrugia, V.~Gakis,
  M.~Hendry, M.~Hohmann, J.~L. Said, J.~Mifsud, and E.~Di~Valentino,
  ``{Teleparallel Gravity: From Theory to Cosmology},''
  \href{http://arxiv.org/abs/2106.13793}{{\tt arXiv:2106.13793 [gr-qc]}}.

\bibitem{Ferraro:2006jd}
R.~Ferraro and F.~Fiorini, ``{Modified teleparallel gravity: Inflation without
  inflaton},'' \href{http://dx.doi.org/10.1103/PhysRevD.75.084031}{{\em Phys.
  Rev.} {\bf D75} (2007)  084031},
\href{http://arxiv.org/abs/gr-qc/0610067}{{\tt arXiv:gr-qc/0610067 [gr-qc]}}.

\bibitem{Ferraro:2008ey}
R.~Ferraro and F.~Fiorini, ``{On Born-Infeld Gravity in Weitzenbock
  spacetime},'' \href{http://dx.doi.org/10.1103/PhysRevD.78.124019}{{\em Phys.
  Rev.} {\bf D78} (2008)  124019},
\href{http://arxiv.org/abs/0812.1981}{{\tt arXiv:0812.1981 [gr-qc]}}.

\bibitem{Bengochea:2008gz}
G.~R. Bengochea and R.~Ferraro, ``{Dark torsion as the cosmic speed-up},''
  \href{http://dx.doi.org/10.1103/PhysRevD.79.124019}{{\em Phys. Rev.} {\bf
  D79} (2009)  124019},
\href{http://arxiv.org/abs/0812.1205}{{\tt arXiv:0812.1205 [astro-ph]}}.

\bibitem{Linder:2010py}
E.~V. Linder, ``{Einstein's Other Gravity and the Acceleration of the
  Universe},'' \href{http://dx.doi.org/10.1103/PhysRevD.81.127301,
  10.1103/PhysRevD.82.109902}{{\em Phys. Rev.} {\bf D81} (2010)  127301},
  \href{http://arxiv.org/abs/1005.3039}{{\tt arXiv:1005.3039 [astro-ph.CO]}}.
[Erratum: Phys. Rev.D82,109902(2010)].

\bibitem{Chen:2010va}
S.-H. Chen, J.~B. Dent, S.~Dutta, and E.~N. Saridakis, ``{Cosmological
  perturbations in f(T) gravity},''
  \href{http://dx.doi.org/10.1103/PhysRevD.83.023508}{{\em Phys. Rev.} {\bf
  D83} (2011)  023508},
\href{http://arxiv.org/abs/1008.1250}{{\tt arXiv:1008.1250 [astro-ph.CO]}}.

\bibitem{Zheng:2010am}
R.~Zheng and Q.-G. Huang, ``{Growth factor in $f(T)$ gravity},''
  \href{http://dx.doi.org/10.1088/1475-7516/2011/03/002}{{\em JCAP} {\bf 1103}
  (2011)  002},
\href{http://arxiv.org/abs/1010.3512}{{\tt arXiv:1010.3512 [gr-qc]}}.

\bibitem{Geng:2011aj}
C.-Q. Geng, C.-C. Lee, E.~N. Saridakis, and Y.-P. Wu, ``{“Teleparallel”
  dark energy},'' \href{http://dx.doi.org/10.1016/j.physletb.2011.09.082}{{\em
  Phys. Lett.} {\bf B704} (2011)  384--387},
\href{http://arxiv.org/abs/1109.1092}{{\tt arXiv:1109.1092 [hep-th]}}.

\bibitem{Bamba:2013jqa}
K.~Bamba, S.~D. Odintsov, and D.~Sáez-Gómez, ``{Conformal symmetry and
  accelerating cosmology in teleparallel gravity},''
  \href{http://dx.doi.org/10.1103/PhysRevD.88.084042}{{\em Phys. Rev.} {\bf
  D88} (2013)  084042},
\href{http://arxiv.org/abs/1308.5789}{{\tt arXiv:1308.5789 [gr-qc]}}.

\bibitem{Kofinas:2014owa}
G.~Kofinas and E.~N. Saridakis, ``{Teleparallel equivalent of Gauss-Bonnet
  gravity and its modifications},''
  \href{http://dx.doi.org/10.1103/PhysRevD.90.084044}{{\em Phys. Rev.} {\bf
  D90} (2014)  084044},
\href{http://arxiv.org/abs/1404.2249}{{\tt arXiv:1404.2249 [gr-qc]}}.

\bibitem{Bahamonde:2019zea}
S.~Bahamonde, K.~Flathmann, and C.~Pfeifer, ``{Photon sphere and perihelion
  shift in weak $f(T)$ gravity},''
  \href{http://dx.doi.org/10.1103/PhysRevD.100.084064}{{\em Phys. Rev. D} {\bf
  100} (2019) no.~8, 084064}, \href{http://arxiv.org/abs/1907.10858}{{\tt
  arXiv:1907.10858 [gr-qc]}}.

\bibitem{Ualikhanova:2019ygl}
U.~Ualikhanova and M.~Hohmann, ``{Parametrized post-Newtonian limit of general
  teleparallel gravity theories},''
  \href{http://dx.doi.org/10.1103/PhysRevD.100.104011}{{\em Phys. Rev. D} {\bf
  100} (2019) no.~10, 104011}, \href{http://arxiv.org/abs/1907.08178}{{\tt
  arXiv:1907.08178 [gr-qc]}}.

\bibitem{DavoodSadatian:2019pvq}
S.~Davood~Sadatian, ``{Effects of viscous content on the modified cosmological
  F(T) model},'' \href{http://dx.doi.org/10.1209/0295-5075/126/30004}{{\em EPL}
  {\bf 126} (2019) no.~3, 30004}.

\bibitem{Bose:2020xdz}
A.~Bose and S.~Chakraborty, ``{Cosmic evolution in f(T) gravity theory},''
  \href{http://dx.doi.org/10.1142/S021773232050296X}{{\em Mod. Phys. Lett. A}
  {\bf 35} (2020) no.~36, 2050296}, \href{http://arxiv.org/abs/2010.16247}{{\tt
  arXiv:2010.16247 [gr-qc]}}.

\bibitem{Zhao:2022gxl}
Y.~Zhao, X.~Ren, A.~Ilyas, E.~N. Saridakis, and Y.-F. Cai, ``{Quasinormal modes
  of black holes in f(T) gravity},''
  \href{http://dx.doi.org/10.1088/1475-7516/2022/10/087}{{\em JCAP} {\bf 10}
  (2022)  087}, \href{http://arxiv.org/abs/2204.11169}{{\tt arXiv:2204.11169
  [gr-qc]}}.

\bibitem{Escamilla-Rivera:2021xql}
C.~Escamilla-Rivera, G.~A. Rave-Franco, and J.~L. Said, ``{f(T, B) Cosmography
  for High Redshifts},'' \href{http://dx.doi.org/10.3390/universe7110441}{{\em
  Universe} {\bf 7} (2021) no.~11, 441},
  \href{http://arxiv.org/abs/2110.05434}{{\tt arXiv:2110.05434 [gr-qc]}}.

\bibitem{Huang:2022slc}
Y.~Huang, J.~Zhang, X.~Ren, E.~N. Saridakis, and Y.-F. Cai, ``{N-body
  simulations, halo mass functions, and halo density profile in f(T)
  gravity},'' \href{http://dx.doi.org/10.1103/PhysRevD.106.064047}{{\em Phys.
  Rev. D} {\bf 106} (2022) no.~6, 064047},
  \href{http://arxiv.org/abs/2204.06845}{{\tt arXiv:2204.06845 [astro-ph.CO]}}.

\bibitem{Blixt:2022rpl}
D.~Blixt, R.~Ferraro, A.~Golovnev, and M.-J. Guzm\'an, ``{Lorentz
  gauge-invariant variables in torsion-based theories of gravity},''
  \href{http://dx.doi.org/10.1103/PhysRevD.105.084029}{{\em Phys. Rev. D} {\bf
  105} (2022) no.~8, 084029}, \href{http://arxiv.org/abs/2201.11102}{{\tt
  arXiv:2201.11102 [gr-qc]}}.

\bibitem{Li:2018ixg}
C.~Li, Y.~Cai, Y.-F. Cai, and E.~N. Saridakis, ``{The effective field theory
  approach of teleparallel gravity, $f(T)$ gravity and beyond},''
  \href{http://dx.doi.org/10.1088/1475-7516/2018/10/001}{{\em JCAP} {\bf 10}
  (2018)  001}, \href{http://arxiv.org/abs/1803.09818}{{\tt arXiv:1803.09818
  [gr-qc]}}.

\bibitem{Casadio:2021zai}
R.~Casadio, I.~Kuntz, and G.~Paci, ``{Quantum fields in teleparallel gravity:
  renormalization at one-loop},''
  \href{http://dx.doi.org/10.1140/epjc/s10052-022-10157-8}{{\em Eur. Phys. J.
  C} {\bf 82} (2022) no.~3, 186}, \href{http://arxiv.org/abs/2110.04325}{{\tt
  arXiv:2110.04325 [hep-th]}}.

\bibitem{Baldazzi:2021kaf}
A.~Baldazzi, O.~Melichev, and R.~Percacci, ``{Metric-Affine Gravity as an
  effective field theory},''
  \href{http://dx.doi.org/10.1016/j.aop.2022.168757}{{\em Annals Phys.} {\bf
  438} (2022)  168757}, \href{http://arxiv.org/abs/2112.10193}{{\tt
  arXiv:2112.10193 [gr-qc]}}.

\bibitem{2008}
S.~Weinberg, ``Effective field theory for inflation,''
  \href{http://dx.doi.org/10.1103/physrevd.77.123541}{{\em Physical Review D}
  {\bf 77} (2008) no.~12, }.
  \url{http://dx.doi.org/10.1103/PhysRevD.77.123541}.

\bibitem{Cheung:2007st}
C.~Cheung, P.~Creminelli, A.~L. Fitzpatrick, J.~Kaplan, and L.~Senatore, ``{The
  Effective Field Theory of Inflation},''
  \href{http://dx.doi.org/10.1088/1126-6708/2008/03/014}{{\em JHEP} {\bf 03}
  (2008)  014}, \href{http://arxiv.org/abs/0709.0293}{{\tt arXiv:0709.0293
  [hep-th]}}.

\bibitem{Cabass:2022avo}
G.~Cabass, M.~M. Ivanov, M.~Lewandowski, M.~Mirbabayi, and M.~Simonovi\'c,
  ``{Snowmass White Paper: Effective Field Theories in Cosmology},'' in {\em
  {2022 Snowmass Summer Study}}.
\newblock 3, 2022.
\newblock \href{http://arxiv.org/abs/2203.08232}{{\tt arXiv:2203.08232
  [astro-ph.CO]}}.

\bibitem{Tsujikawa:2014mba}
S.~Tsujikawa, ``{The effective field theory of inflation/dark energy and the
  Horndeski theory},''
  \href{http://dx.doi.org/10.1007/978-3-319-10070-8_4}{{\em Lect. Notes Phys.}
  {\bf 892} (2015)  97--136}, \href{http://arxiv.org/abs/1404.2684}{{\tt
  arXiv:1404.2684 [gr-qc]}}.

\bibitem{Solomon}
A.~R. Solomon and M.~Trodden, ``{Higher-derivative operators and effective
  field theory for general scalar-tensor theories},''
  \href{http://dx.doi.org/10.1088/1475-7516/2018/02/031}{{\em JCAP} {\bf 02}
  (2018)  031}, \href{http://arxiv.org/abs/1709.09695}{{\tt arXiv:1709.09695
  [hep-th]}}.

\bibitem{Maldacena:2011nz}
J.~M. Maldacena and G.~L. Pimentel, ``{On graviton non-Gaussianities during
  inflation},'' \href{http://dx.doi.org/10.1007/JHEP09(2011)045}{{\em JHEP}
  {\bf 09} (2011)  045}, \href{http://arxiv.org/abs/1104.2846}{{\tt
  arXiv:1104.2846 [hep-th]}}.

\bibitem{Crisostomi:2017ugk}
M.~Crisostomi, K.~Noui, C.~Charmousis, and D.~Langlois, ``{Beyond Lovelock
  gravity: Higher derivative metric theories},''
  \href{http://dx.doi.org/10.1103/PhysRevD.97.044034}{{\em Phys. Rev. D} {\bf
  97} (2018) no.~4, 044034}, \href{http://arxiv.org/abs/1710.04531}{{\tt
  arXiv:1710.04531 [hep-th]}}.

\bibitem{Mylova:2019jrj}
M.~Mylova, ``{Chiral primordial gravitational waves in extended theories of
  Scalar-Tensor gravity},'' \href{http://arxiv.org/abs/1912.00800}{{\tt
  arXiv:1912.00800 [gr-qc]}}.

\bibitem{Mukohyama:2022enj}
S.~Mukohyama and V.~Yingcharoenrat, ``{Effective field theory of black hole
  perturbations with timelike scalar profile: formulation},''
  \href{http://dx.doi.org/10.1088/1475-7516/2022/09/010}{{\em JCAP} {\bf 09}
  (2022)  010}, \href{http://arxiv.org/abs/2204.00228}{{\tt arXiv:2204.00228
  [hep-th]}}.

\bibitem{Hui:2021cpm}
L.~Hui, A.~Podo, L.~Santoni, and E.~Trincherini, ``{Effective Field Theory for
  the perturbations of a slowly rotating black hole},''
  \href{http://dx.doi.org/10.1007/JHEP12(2021)183}{{\em JHEP} {\bf 12} (2021)
  183}, \href{http://arxiv.org/abs/2111.02072}{{\tt arXiv:2111.02072
  [hep-th]}}.

\bibitem{petrov}
A.~A. Petrov and A.~E. Blechman, {\em Effective field theories}.
\newblock World Scientific, 2015.

\bibitem{baumann2015inflation}
D.~Baumann and L.~McAllister, {\em Inflation and string theory}.
\newblock Cambridge University Press, 2015.

\bibitem{burgess2020introduction}
C.~P. Burgess, {\em Introduction to effective field theory}.
\newblock Cambridge University Press, 2020.

\bibitem{Planck:2018jri}
{\bf Planck} Collaboration, Y.~Akrami {\em et al.}, ``{Planck 2018 results. X.
  Constraints on inflation},''
  \href{http://dx.doi.org/10.1051/0004-6361/201833887}{{\em Astron. Astrophys.}
  {\bf 641} (2020)  A10}, \href{http://arxiv.org/abs/1807.06211}{{\tt
  arXiv:1807.06211 [astro-ph.CO]}}.

\bibitem{ostrogradsky1850memoire}
M.~Ostrogradsky, ``{M\'emoires sur les \'equations diff\'erentielles, relatives
  au probl\`eme des isop\'erim\`etres},'' {\em Mem. Acad. St. Petersbourg} {\bf
  6} (1850) no.~4, 385--517.

\bibitem{Woodard:2015zca}
R.~P. Woodard, ``{Ostrogradsky's theorem on Hamiltonian instability},''
  \href{http://dx.doi.org/10.4249/scholarpedia.32243}{{\em Scholarpedia} {\bf
  10} (2015) no.~8, 32243}, \href{http://arxiv.org/abs/1506.02210}{{\tt
  arXiv:1506.02210 [hep-th]}}.

\bibitem{georgi1991shell}
H.~Georgi, ``On-shell effective field theory,'' {\em Nuclear Physics B} {\bf
  361} (1991) no.~2, 339--350.

\bibitem{Grosse-Knetter:1993tae}
C.~Grosse-Knetter, ``{Effective Lagrangians with higher derivatives and
  equations of motion},''
  \href{http://dx.doi.org/10.1103/PhysRevD.49.6709}{{\em Phys. Rev. D} {\bf 49}
  (1994)  6709--6719}, \href{http://arxiv.org/abs/hep-ph/9306321}{{\tt
  arXiv:hep-ph/9306321}}.

\bibitem{Arzt:1993gz}
C.~Arzt, ``{Reduced effective Lagrangians},''
  \href{http://dx.doi.org/10.1016/0370-2693(94)01419-D}{{\em Phys. Lett. B}
  {\bf 342} (1995)  189--195}, \href{http://arxiv.org/abs/hep-ph/9304230}{{\tt
  arXiv:hep-ph/9304230}}.

\bibitem{Manohar:2018aog}
A.~V. Manohar, ``{Introduction to Effective Field Theories},''
  \href{http://arxiv.org/abs/1804.05863}{{\tt arXiv:1804.05863 [hep-ph]}}.

\bibitem{Weitzenbock1923}
R.~Weitzenb\"{o}ock, {\em `Invariantentheorie'}.
\newblock Noordhoff, Gronningen, 1923.

\bibitem{Hayashi:1979qx}
K.~Hayashi and T.~Shirafuji, ``{New General Relativity},''
  \href{http://dx.doi.org/10.1103/PhysRevD.19.3524}{{\em Phys. Rev. D} {\bf 19}
  (1979)  3524--3553}. [Addendum: Phys.Rev.D 24, 3312--3314 (1982)].

\bibitem{Bahamonde:2015zma}
S.~Bahamonde, C.~G. Böhmer, and M.~Wright, ``{Modified teleparallel theories
  of gravity},'' \href{http://dx.doi.org/10.1103/PhysRevD.92.104042}{{\em Phys.
  Rev.} {\bf D92} (2015) no.~10, 104042},
\href{http://arxiv.org/abs/1508.05120}{{\tt arXiv:1508.05120 [gr-qc]}}.

\bibitem{Farrugia:2016qqe}
G.~Farrugia and J.~L. Said, ``{Stability of the flat FLRW metric in $f(T)$
  gravity},'' \href{http://dx.doi.org/10.1103/PhysRevD.94.124054}{{\em Phys.
  Rev.} {\bf D94} (2016) no.~12, 124054},
\href{http://arxiv.org/abs/1701.00134}{{\tt arXiv:1701.00134 [gr-qc]}}.

\bibitem{Zumalacarregui:2013pma}
M.~Zumalac\'arregui and J.~Garc\'\i{}a-Bellido, ``{Transforming gravity: from
  derivative couplings to matter to second-order scalar-tensor theories beyond
  the Horndeski Lagrangian},''
  \href{http://dx.doi.org/10.1103/PhysRevD.89.064046}{{\em Phys. Rev. D} {\bf
  89} (2014)  064046}, \href{http://arxiv.org/abs/1308.4685}{{\tt
  arXiv:1308.4685 [gr-qc]}}.

\bibitem{Bahamonde:2019shr}
S.~Bahamonde, K.~F. Dialektopoulos, and J.~Levi~Said, ``{Can Horndeski Theory
  be recast using Teleparallel Gravity?},''
  \href{http://dx.doi.org/10.1103/PhysRevD.100.064018}{{\em Phys. Rev. D} {\bf
  100} (2019) no.~6, 064018}, \href{http://arxiv.org/abs/1904.10791}{{\tt
  arXiv:1904.10791 [gr-qc]}}.

\bibitem{Bahamonde:2019ipm}
S.~Bahamonde, K.~F. Dialektopoulos, V.~Gakis, and J.~Levi~Said, ``{Reviving
  Horndeski theory using teleparallel gravity after GW170817},''
  \href{http://dx.doi.org/10.1103/PhysRevD.101.084060}{{\em Phys. Rev. D} {\bf
  101} (2020) no.~8, 084060}, \href{http://arxiv.org/abs/1907.10057}{{\tt
  arXiv:1907.10057 [gr-qc]}}.

\bibitem{Hohmann:2020zre}
M.~Hohmann, ``{Complete classification of cosmological teleparallel
  geometries},'' \href{http://dx.doi.org/10.1142/S0219887821400053}{{\em Int.
  J. Geom. Meth. Mod. Phys.} {\bf 18} (2021) no.~supp01, 2140005},
  \href{http://arxiv.org/abs/2008.12186}{{\tt arXiv:2008.12186 [gr-qc]}}.

\end{thebibliography}\endgroup

\end{document}